\documentclass[%
]{amsart}

\usepackage{amsmath,amssymb}
\usepackage{graphicx}
\usepackage{bm}
\usepackage{hyperref}

\usepackage{subfigure}
\usepackage[font=footnotesize]{caption}
 \usepackage{url}
 
\usepackage{tikz}
\usetikzlibrary{arrows,shapes,calc,positioning}

  \tikzstyle{resident}=[circle,fill=blue,draw=none,text=blue]
  \tikzstyle{mutant}=[resident,fill=red,text=red]
  \tikzstyle{reproductor}=[circle,very thick, fill=red,draw=green!60!black,text=green]
  \tikzstyle{dead}=[circle,very thick,fill=blue,draw=black,text=black]
  \definecolor{zzzzzz}{rgb}{0.4,0.4,0.4}
  \definecolor{ffttqq}{rgb}{1,0.2,0}

\let\rs\setminus

\newcommand{\num}[1]{\lvert #1 \rvert}
\newcommand{\norm}[1]{\lVert #1 \rVert_\infty}

\usepackage[T1]{fontenc}
\usepackage{listings}
\usepackage[scaled=.8]{beramono}
\usepackage{navigator}
\usepackage{fullpage}
\usepackage[warn]{textcomp}

\usepackage{geometry}

\usepackage{pdfpages}

\begin{document}

\newgeometry{left=1.75in,right=1.75in,footskip=30pt}

\title{Suppressors of Selection}

%

\author[F. Alcalde]{Fernando Alcalde Cuesta$^{1,2}$}
\address{$^1$~GeoDynApp - ECSING Group, Spain. \vspace*{-2ex}
}
\address{$^2$~Departamento de Matem\'aticas, Universidade de Santiago de Compostela, E-15782, Santiago de Compostela, Spain. }
\email{fernando.alcalde@usc.es}

\author[P. G. Sequeiros]{Pablo Gonz\'alez Sequeiros$^{1,3}$}
\address{$^1$~GeoDynApp - ECSING Group  \vspace*{-2ex}
}
\address{$^3$~Departamento de Did\'acticas Aplicadas, Facultade de Formaci\'on do Profesorado, Universidade de Santiago de Compostela, Avda. Ram\'on Ferreiro 10, E-27002 Lugo, Spain}
\email{pablo.gonzalez.sequeiros@usc.es}%

\author[\'A. Lozano]{\'Alvaro Lozano Rojo$^{1,4,5}$}
\address{$^1$~GeoDynApp - ECSING Group  \vspace*{-2ex}
}
\address{$^4$~Centro Universitario de la Defensa, Academia General Militar, Ctra. Huesca s/n. E-50090 Zaragoza, Spain \vspace*{-2ex}
}%
\address{$^5$~Instituto Universitario de Matem\'aticas y Aplicaciones, Universidad de Zaragoza, Spain}
\email{alvarolozano@unizar.es}

\date{\today}

\begin{abstract}
Inspired by recent works on evolutionary graph theory, an area of growing interest in mathematical and computational biology, we present the first known examples of undirected structures acting as suppressors of selection for any fitness value $r > 1$.
 This means that the average fixation probability of an advantageous mutant or invader individual placed at some node is strictly less than that of this individual placed in a well-mixed population. This leads the way to study more robust structures less prone to invasion, contrary to what happens with the amplifiers of selection where the fixation probability is increased on average for advantageous invader individuals. A few families of amplifiers are known, although some effort was required to prove it. Here, we use computer aided techniques  to find an exact analytical expression of the fixation probability for some graphs of small order (equal to $6$, $8$ and $10$) proving that selection is effectively reduced for $r > 1$. Some numerical experiments using Monte Carlo methods are also performed for larger graphs.

\end{abstract}

\maketitle


\section{Introduction}
Evolutionary dynamics has been classically studied for well-mixed populations, but there is a wide interest in the evolution of complex networks after site invasion. The process transforming nodes occupied by residents into nodes occupied by mutants or invaders is described by the \emph{Moran model}. Introduced by Moran \cite{M} as the Markov chain counting the number of invading mutants in a well-mixed population, it was adapted to weighted graphs by Lieberman et al. \cite{LHN} and Nowak \cite{N} (see also~\cite{BR, BHRS, SR2, ShakarianBio, Voorhees1}). For undirected networks where links have no orientation, invaders will either become extinct or take over the whole population, reaching one of the two absorbing 
states, \emph{extinction} or \emph{fixation}. The \emph{fixation probability} is the fundamental quantity in the stochastic evolutionary analysis of a finite population. 

If the population is well-mixed, at the beginning, one single node is chosen to be occupied by an invader individual among a population of $N$ resident individuals. Afterwards, an individual is randomly chosen for reproduction, with probability proportional to its reproductive advantage ($1$ for residents and $r \geq 1$ for invaders), and its clonal offspring replaces another individual chosen at random. In this case, the fixation probability is given by 
\begin{equation} \label{eq:FPmin}
  \Phi_0(r)  = \frac{1- r^{-1}}{1 - r^{-N}} = \frac{r^{N-1}}{r^{N-1} + r^{N-2} + \dots + r + 1}.
\end{equation}

In evolutionary network theory, the nodes are occupied by resident or invader individuals and the replacements are limited to the nodes which are connected by oriented links. According to the Circulation Theorem \cite{LHN}, any weight-balanced network has the same fixation probability as the well-mixed population of the same size $N$. In the undirected case, this means that the \emph{temperature} 
$T_i = \sum_{j \sim i} 1/d_j$ of every vertex $i$ (where $j$ is a neighbor of $i$ and $d_j$ is the number 
of neighbors of $j$) is constant, and the network is said to be \emph{isothermal}. But there are graph structures altering substantially the behavior of the fixation probability depending on the fitness. For example, the (average) fixation probability in the oriented line is equal to $1/N$ and the reproductive advantage of the invader individuals is completely suppressed. But in the directed case, absorbing barriers may not be accessible from any state, and the fixation probability may be even null (see \cite{AGSL} for an example). 

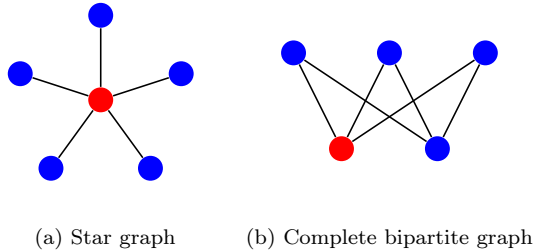
\begin{figure}
  \subfigure[~Star graph]{
    \begin{tikzpicture}[scale=0.75]
   
    \path[use as bounding box] (-2,-2) rectangle (2,1.75);
    
    \node[resident] (A)  at (90:1.5)  {};
    \foreach \a/\b in {1/B, 2/C, 3/D, 4/E}{
      \node[resident] (\b) at (90+72*\a:1.5) {};
    }
    
    \node[mutant] (O) at (0,0) {};
  
    \foreach \to in {A, B, C, D, E}
      \draw [semithick] (O) -- (\to);
  
  \end{tikzpicture}
    \label{fig:star}
  }
  \subfigure[~Complete bipartite graph]{
    \begin{tikzpicture}[scale=0.85]
      \path[use as bounding box] (-2.2,-2.5) rectangle (2.2,0);
   
      \node[resident]   (A) at ( 0   , 0  ) {};
      \node[resident] (B) at (-1.5 , 0  ) {};
      \node[resident] (C) at ( 1.5 , 0  ) {};
      \node[resident] (D) at ( 0.75,-1.5) {};
      \node[mutant] (E) at (-0.75,-1.5) {};
  
      \foreach \from/\to in {A/D, A/E, B/D, B/E, C/D, C/E}
        \draw [semithick] (\from) -- (\to);
      
    \end{tikzpicture}
    \label{fig:bipartite}
  }
  \caption{{\bf Star and complete bipartite graphs.} 
  (a) In the star graph $K_{1,m}$, the center is connected with $m$ peripheral nodes. 
  (b) The vertex set of a complete bipartite graph $K_{n,m}$ is divided into two disjoint sets interconnected by edges.}
  \label{fig1}
\end{figure}

Thus, we focus our attention on undirected networks where absorbing barriers can be reached from any state. As showed in \cite{LHN,N} (see also
\cite{Adlam&al}), there are directed and undirected graph structures that asymptotically amplify this advantage. The fixation probability of a complete bipartite network $K_{N-m,m}$ (described in  Figure~\ref{fig1}) converges to the same limit as the fixation probability
\begin{equation} \label{eq:FPmax}
  \Phi_2(r) = \Phi_0(r^2)= \frac{1- r^{-2}}{1 - r^{-2N}} 
\end{equation}
of the Moran process with fitness $r^2$ as $m \to \infty$ and $N-m$ is constant \cite{AGSL}. Assuming that fitness differences are amplified or reduced for network sequences of increasing size, a notion of \emph{amplifier} and \emph{suppressor of selection} has been introduced in \cite{Adlam&al} for several initialization types (describing the initial distribution of the invasion process). To distinguish both dynamics, a numerical analysis for a few fitness values $r=0.75,1,1.25, 1.5, 1.75$ has been done in \cite{HindersinTraulsen} for birth-death and death-birth processes on 
directed and undirected graphs (see \cite{KKK} for a comparative analysis of both update mechanisms).

Here we always assume that the distribution is uniform: the probability that a node will be occupied by the initial invader is equal for all the nodes (see Eq~\eqref{eq:average}). We say that a network is an \emph{amplifier of selection} if the fixation probability function 
$\Phi(r) > \Phi_0(r)$ and a \emph{suppressor of selection} if $\Phi(r) < \Phi_0(r)$ for all $r >1$. Notice that $\Phi(1) = 1/N$ and the inequalities 
must be reversed  for $r<1$. Due to the exact analytical expression given by Monk et al. \cite{Monk&al} using martingales (see also \cite{Adlam&al}), one can see that star graphs and complete bipartite graph are amplifiers of natural selection whose fixation functions are bounded from above by $\Phi_2(r)$.
One could also ask if the fixation function is always greater than or equal to that of a well-mixed population of the same size, denoted by $\Phi_0(r)$, at least from some fitness value. A negative answer to the first question was given in \cite{BRS2} for fitness values $r \leq 10$. The aim of the paper is to prove that both questions are not true, exhibiting examples of graphs with $6$, $8$ and $10$ vertices which are suppressors of selection for any fitness value $r > 1$. From the point of view of robustness against invasion \cite{SciRep}, these graphs are more robust than complete graphs (being now necessary to add a sign to $\norm{\Phi-\Phi_0} = \sup_{r \geq 1} \lvert\Phi(r)-\Phi_0(r)\rvert$). 
Better yet,  we propose a complete family of graphs of even order $2n+2$ with $n \geq 2$, called \emph{$\ell$-graphs}, which we believe are suppressors of selection. The proof of this assertion for the graphs of order $6$, $8$ and $10$ is completed with a numerical simulation for larger orders.  Some other variants are also explored numerically in order to understand why they are suppressors of selection.

\section{Results}

All the examples of so-called suppressors of selection given in \cite{LHN,N} are directed graphs. The abundance of amplifiers and suppressors of selection has been explored numerically by Hindersin et al. in \cite{HindersinTraulsen} for this kind of graphs under birth-death and death-birth updating. Different types of initialization or placement of new invaders have been distinguished in \cite{Adlam&al} in order to classify different evolutionary dynamics on directed graphs. As explained, we focus our attention on undirected  graphs under uniform initialization. 

\begin{figure*}
  \subfigure[~Graph $l_6$]{
    \begin{tikzpicture}[scale=0.7]
   
    \path[use as bounding box] (-1.8,-2) rectangle (2.5,2);
    
    \node[resident] (A)  at (45:1.5)  {};
    \foreach \a/\b in {1/B, 2/C, 3/D}{
      \node[resident] (\b) at (45+90*\a:1.5) {};
    }
    
    \node[resident] (e-) at (-2,0) {};
       \node[resident] (e+) at (2,0) {};
       
       \draw [semithick] (e-) to [bend left=45] (0,2.5) to [bend left=45] (e+);
   
  \foreach \from in {A, B, C, D}
    \foreach \to in {A, B, C, D}
       \draw [semithick] (\from) -- (\to);
     \foreach \+ in {A, D}
           \draw [semithick] (\+) -- (e+);
       \foreach \- in {B, C}
      \draw [semithick] (\-) -- (e-);

  \end{tikzpicture}
    \label{fig:l6}
  }
  \subfigure[~Graph $l_8$]{
   \begin{tikzpicture}[scale=0.7]
   
    \path[use as bounding box] (-3,-2) rectangle (3,2);
    
    \node[resident] (A)  at (0:1.5)  {};
    \foreach \a/\b in {1/B, 2/C, 3/D, 4/E, 5/F}{
      \node[resident] (\b) at (0+60*\a:1.5) {};
    }
    
    \node[resident] (e-) at (-2.7,0) {};
       \node[resident] (e+) at (2.7,0) {};
       
       \draw [semithick] (e-) to [bend left=45] (0,2.7) to [bend left=45] (e+);
   
  \foreach \from in {A, B, C, D, E, F}
    \foreach \to in {A, B, C, D, E, F}
       \draw [semithick] (\from) -- (\to);
     \foreach \+ in {A,B, F}
           \draw [semithick] (\+) -- (e+);
       \foreach \- in {C, D, E}
      \draw [semithick] (\-) -- (e-);

  \end{tikzpicture}
    \label{fig:l8}
  }
   \quad
  \subfigure[~Graph $l_{10}$]{
   \begin{tikzpicture}[scale=0.7]
   
    \path[use as bounding box] (-2.8,-2) rectangle (2.8,3.2);
    
    \node[resident] (A)  at (22.5:1.4)  {};
    \foreach \a/\b in {1/B, 2/C, 3/D, 4/E, 5/F, 6/G, 7/H}{
      \node[resident] (\b) at (22.5+45*\a:1.4) {};
    }
    
    \node[resident] (e-) at (-2.8,0) {};
       \node[resident] (e+) at (2.8,0) {};
       
       \draw [semithick] (e-) to [bend left=45] (0,2.9) to [bend left=45] (e+);
   
  \foreach \from in {A, B, C, D, E, F, G, H}
    \foreach \to in {A, B, C, D, E,  F, G, H}
       \draw [semithick] (\from) -- (\to);
           \draw [semithick] (A) -- (e+);
             \draw [semithick] (H) -- (e+);
                \draw [semithick] (B) to [bend left=17] (e+);
                 \draw [semithick] (G) to [bend right=17] (e+);
	 \draw [semithick] (D) -- (e-);
             \draw [semithick] (E) -- (e-);
                \draw [semithick] (F) to [bend left=17] (e-);
                 \draw [semithick] (C) to [bend right=17] (e-);

  \end{tikzpicture}
    \label{fig:l10}
  }
  \caption{Graphs of size $6$, $8$ and $10$ in the $l$-family.}
  \label{fig2}
\end{figure*}
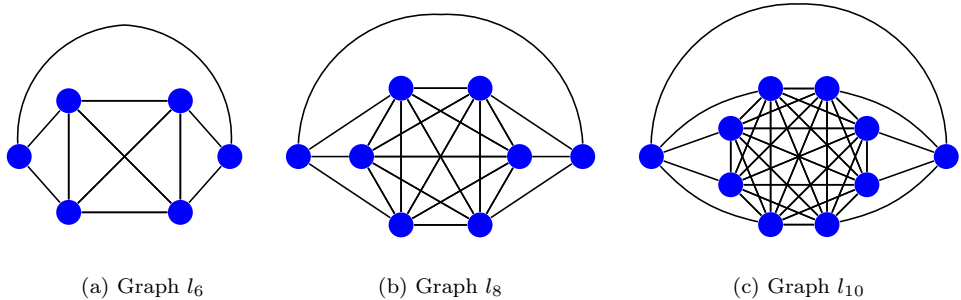

Firstly, working with the \emph{FinisTerrae2} supercomputer (we used 1024 cores of Haskell 2680v3 CPUs for almost 3 days) installed at CESGA,
we computed the fixation probability of all undirected graphs of order $10$ or less
for fitness values $r$ varying from $0.25$ to $10$ with step size of $0.25$ \cite{AGSL2}. We found an unique suppressor of selection of order $6$, namely the graph $\ell_6$, although there are other possible suppressors in orders varying from $7$ to $10$. We constructed the graphs $\ell_8$ and $\ell_{10}$ (as well the whole $\ell$-family) from this initial example. More precisely, we call \emph{$\ell$-graph} an undirected graph of even order $N=2n+2 \geq 6$ obtained from the complete graph $K_{2n}$ by dividing its vertex set into two halves with $n \geq 2$ vertices and adding $2$ extra vertices. Each of them is connected to one of the halves of $K_{2n}$ and with the other extra vertex.
Graphs $\ell_6$, $\ell_8$ and $\ell_{10}$ of order $6$, $8$ and $10$ are shown in Figure~\ref{fig2}. 

\subsection{Graphs $\ell_6$, $\ell_8$ and $\ell_{10}$ are suppressors of selection}
Computer aided techniques has been used to find exact analytical expressions of the fixation function $\Phi$ for the first elements of this family with orders 
$6$, $8$ and $10$ (see Figure~\ref{fig2}). This computation proves that the graphs $\ell_6$, $\ell_8$ and $\ell_{10}$ are suppressors of selection with fixation functions $\Phi(r) < \Phi_0(r)$ for all $r > 1$ and $\Phi(r) > \Phi_0(r)$ for all $r < 1$. 

At first, to bound the fixation probability from above, one could try to stop the process on $\ell_{2n+2}$ at the time that some extra vertex is occupied by a mutant. But as we will see later, the evolution from that time on seems play an essential role in determining the suppressor character of the graph. 
Like for star and looping star graphs, which are amplifiers of selection for uniform initialization \cite{Adlam&al}, we needed  then to find the exact analytical expression of the fixation probability. Unfortunately, the elegant martingale method (which is based on Doob's optional stopping theorem \cite{KT}) proposed in \cite{Monk&al} and used in \cite{Adlam&al} is not useful for $\ell_6$, $\ell_8$ and $\ell_{10}$. We have had to implement a specific method to compute exactly their fixation probability.

As we shall see in the description of the mathematical model in the Methods section, 
the fixation function $\Phi(r)$ is always a rational function given as the quotient of two rational polynomials $\Phi'(r)$ and $\Phi''(r)$ of degree bounded above by $2^N-2$. Using the symmetries of each $\ell$-graph, we can lower this bound to a quantity 
\begin{equation} \label{eq:d}
d = \frac{N(N+1)}{2} -2 \ll 2^N-2
\end{equation}
as proved in the Methods section, and hence there are and hence there are only $2(d+1)$ coefficients involved in $\Phi(r)$. Since $\Phi(r)$ converges to $1$ as $r \to+\infty$, the leading coefficients of $\Phi'(r)$ and $\Phi''(r)$ can be assumed to be $1$ and that number is reduced to $2d$. Then we can replace the system of $2^N$ linear equations defining the fixation function $\Phi(r)$ (see Eq~\eqref{eq:fixation}) with a system of $2d$ linear equations (see Eq~\eqref{eq:reduced}) corresponding to the $2d$ rational coefficients of $\Phi'(r)$ and $\Phi''(r)$, which arise from evaluating $\Phi(r)$ for
integer and rational values of the fitness $r$ varying from $1$ to $d+1$ and  from $1/2$ to $1/d$.
Finally, we wrote a SageMath program \cite{sage} (which is added to the Supporting Information) to compute the exact fixation probability $\Phi(r)$ of the graphs $\ell_6$, $\ell_8$ and $\ell_{10}$ for these fitness values and to solve the reduced linear system. 
Once the fixation function $\Phi$ has been calculated, the sign of the difference $\Delta = \Phi - \Phi_0$ is analyzed to confirm that $\Delta(r) < 0$ for all $r >  1$. 
%
In the Methods section, we give a more detailed explanation of both theoretical and computational arguments used to have exact analytical expressions of the fixation function $\Phi$ for $\ell_6$, $\ell_8$ and $\ell_{10}$. The exact values of $\Phi$ and $\Delta$ are given in the Supporting Information of the paper. 

\subsection{Numerical experiments in larger orders. Further examples}

\begin{figure}[!t]
\centering
\includegraphics[width=3.2in]{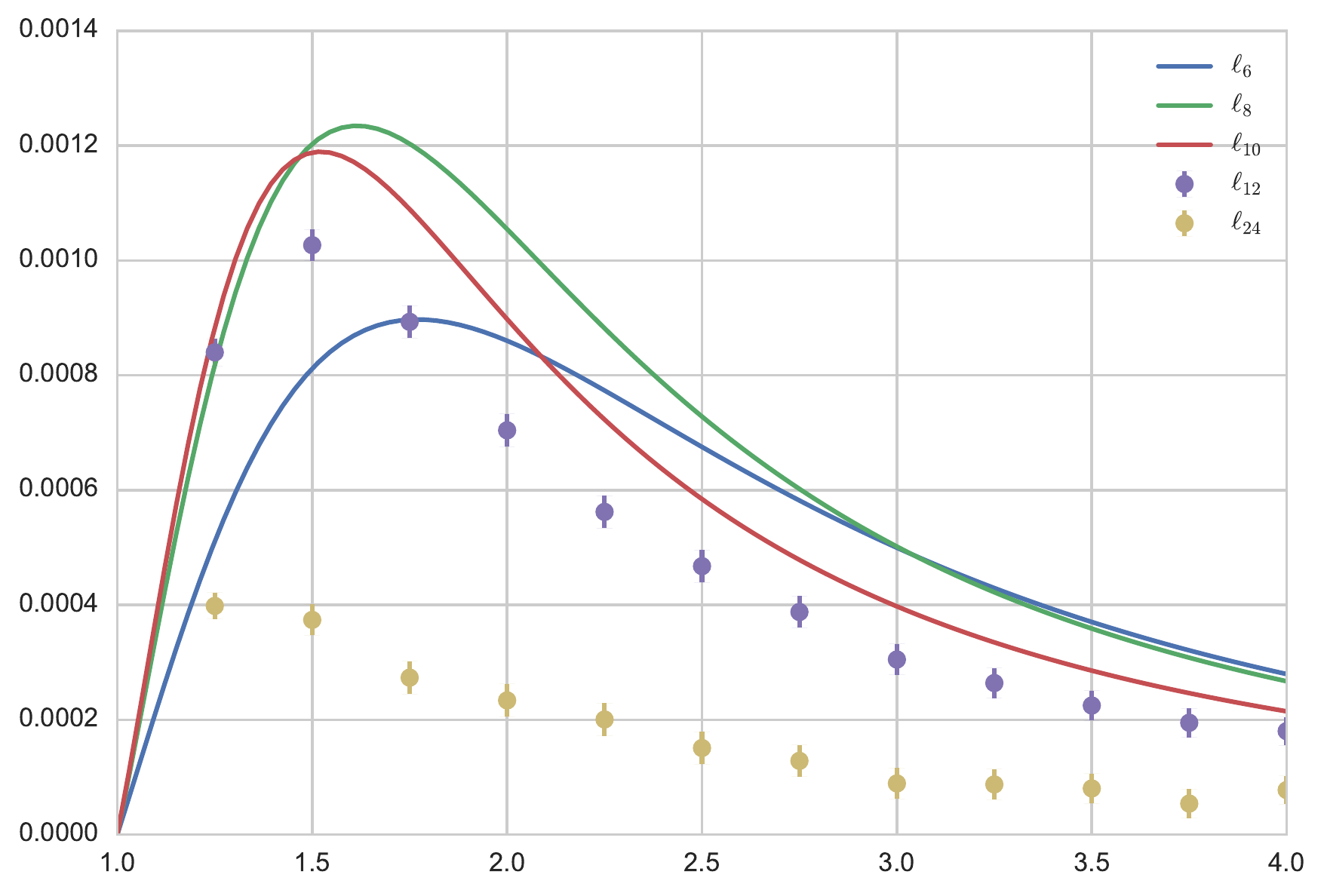}%
 \caption{{\bf The  exact differences $\Phi_0(r) - \Phi(r)$ for $\ell_6$, $\ell_8$ and $\ell_{10}$ and some estimates for $\ell_{12}$ and $\ell_{24}$.} The functions 
 $\Phi_0(r) - \Phi(r)$ associated to the $\ell$-graphs of order $6$, $8$, and $10$ are represented for fitness values $r$ varying from $1$ to $4$. For the $\ell$-graphs of order $12$ and $24$, we applied the Monte Carlo method to compute the difference between the fixation probabilities of each graph and the complete graph of the same order using $10^8$ trials for each fitness value $r$ varying from $0$ to $4$ with step size of $0.25$. Notice that $\Phi_0(r) - \Phi(r)$ converges to $0$ as the order of the graph goes to infinity for all fitness value $0 < r < +\infty$.}
  \label{fig3}
\end{figure}

 However, 
the method used for  for $\ell_6$, $\ell_8$ and $\ell_{10}$  does not seem to be applicable to larger orders since it would require a substantial amount of memory and computation time. Therefore, we explored the suppression of selection for other graphs in the $\ell$-family using Monte Carlo simulation (applying the \emph{Loop-Erasing} technique of \cite{AGSL} to speedup the computations). Even if it does not require much memory and can be parallelized on a computer cluster, a very large number of trials ---namely $10^8$ trails for each fitness value--- has been necessary to compare the fixation probability of 
$\ell_{12}$ and $\ell_{24}$ with that of the complete graphs of the same order. In fact, since the fixation function of the $\ell$-graph of order $N$ approaches the fixation function $\Phi_0(r)$ given by Eq~\eqref{eq:FPmin}, we should need to increase this number more and more as $N$ goes to $\infty$. Anyway, for fitness values $r$ varying from $0$ to $4$ with step size of $0.25$, we showed that the $\ell$-graphs of orders $12$ and $24$ are also suppressors of selection as can be seen in Figure~\ref{fig3}.

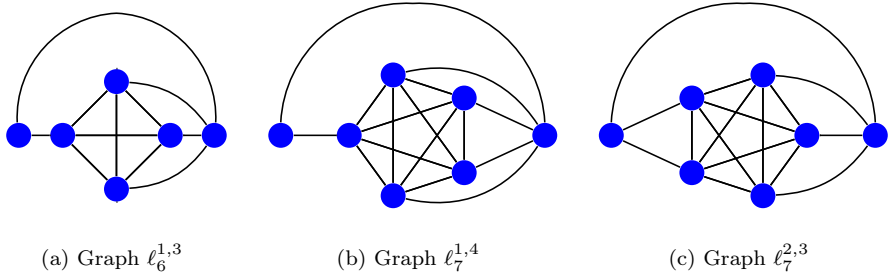
\begin{figure}[!t]
  \subfigure[~Graph $\ell_6^{1,3}$]{
    \begin{tikzpicture}[scale=0.65]
   
    \path[use as bounding box] (-2.5,-2) rectangle (2.5,1.5);
    
    \node[resident] (A)  at (0:1.1)  {};
    \foreach \a/\b in {1/B, 2/C, 3/D}{
      \node[resident] (\b) at (90*\a:1.1) {};
    }
    
    \node[resident] (e-) at (-2,0) {};
       \node[resident] (e+) at (2,0) {};
       
       \draw [semithick] (e-) to [bend left=45] (0,2.5) to [bend left=45] (e+);
   
  \foreach \from in {A, B, C, D}
    \foreach \to in {A, B, C, D}
       \draw [semithick] (\from) -- (\to);
       \draw [semithick] (A) -- (e+);
            \draw [semithick] (B) to [bend left=25] (e+);
            \draw [semithick] (D) to [bend right=25] (e+);
       \foreach \- in {C}
      \draw [semithick] (\-) -- (e-);

  \end{tikzpicture}}
      \label{fig:l613} 
  \subfigure[~Graph $\ell_7^{1,4}$]{
   \begin{tikzpicture}[scale=0.65]
   
    \path[use as bounding box] (-3.2,-2) rectangle (3.2,2);
    
    \node[resident] (A)  at (36:1.3)  {};
    \foreach \a/\b in {1/B, 2/C, 3/D, 4/E}{
      \node[resident] (\b) at (36+72*\a:1.3) {};
    }
    
    \node[resident] (e-) at (-2.7,0) {};
       \node[resident] (e+) at (2.7,0) {};
       
       \draw [semithick] (e-) to [bend left=45] (0,2.7) to [bend left=45] (e+);
   
  \foreach \from in {A, B, C, D, E}
    \foreach \to in {A, B, C, D, E}
       \draw [semithick] (\from) -- (\to);
     \foreach \+ in {A,E}
           \draw [semithick] (\+) -- (e+);
            \draw [semithick] (B) to [bend left=35] (e+);
            \draw [semithick] (D) to [bend right=35] (e+);
       \foreach \- in { C}
      \draw [semithick] (\-) -- (e-);

  \end{tikzpicture}}
    \label{fig:l714}
  \subfigure[~Graph $\ell_{7}^{2,3}$]{
    \begin{tikzpicture}[scale=0.65]
   
    \path[use as bounding box] (-3.2,-2) rectangle (3.2,2.5);
    
    \node[resident] (A)  at (0:1.3)  {};
    \foreach \a/\b in {1/B, 2/C, 3/D, 4/E}{
      \node[resident] (\b) at (72*\a:1.3) {};
    }
    
    \node[resident] (e-) at (-2.7,0) {};
       \node[resident] (e+) at (2.7,0) {};
       
       \draw [semithick] (e-) to [bend left=45] (0,2.7) to [bend left=45] (e+);
   
  \foreach \from in {A, B, C, D, E}
    \foreach \to in {A, B, C, D, E}
       \draw [semithick] (\from) -- (\to);
           \draw [semithick] (A) -- (e+);
            \draw [semithick] (B) to [bend left=25] (e+);
            \draw [semithick] (E) to [bend right=25] (e+);
       \foreach \- in { C,D}
      \draw [semithick] (\-) -- (e-);

  \end{tikzpicture}}
    \label{fig:l723}
  \caption{{\bf Unbalanced $\ell$-graphs of order $6$ and $7$.}}
  \label{unbalanced}
\end{figure}

\begin{figure}[!t]
\centering
\includegraphics[width=3.5in]{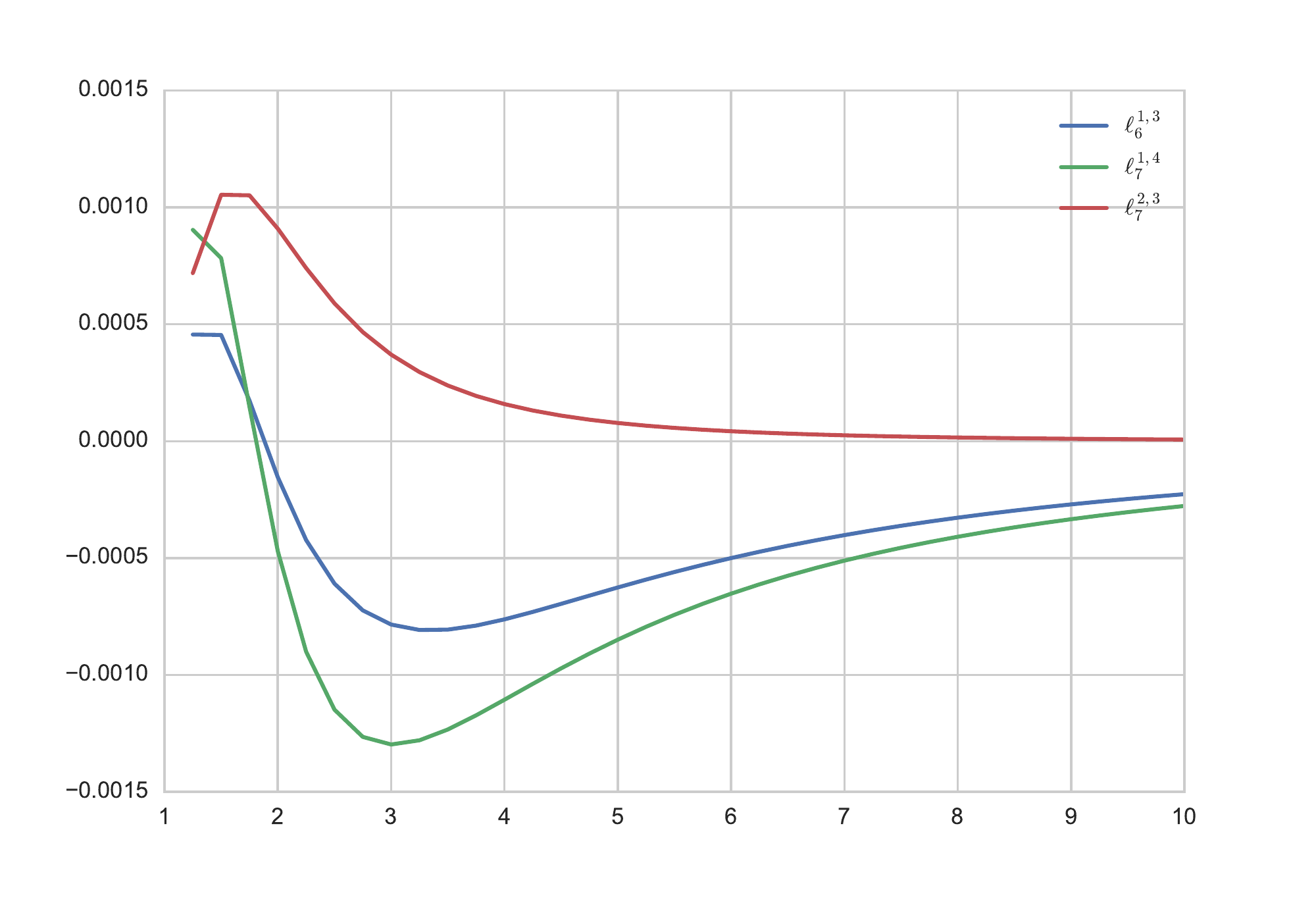}%
 \caption{{\bf The differences $\Delta(r) = \Phi(r) - \Phi_0(r)$ for the unbalanced graphs $\ell_6^{1,3}$,  $\ell_7^{1,4}$, and $\ell_7^{2,3}$.} The differences $\Delta(r) = \Phi(r) - \Phi_0(r)$ have been estimated using $10^8$ trials for each fitness value $r$ varying from $0$ to $10$ with step size of $0.25$. }
  \label{fig5}
\end{figure}

\begin{figure}[!t]
\centering
\vspace{-3ex}
\subfigure[Unbalanced $\ell$-graphs of order $8$]{\includegraphics[width=3.2in]{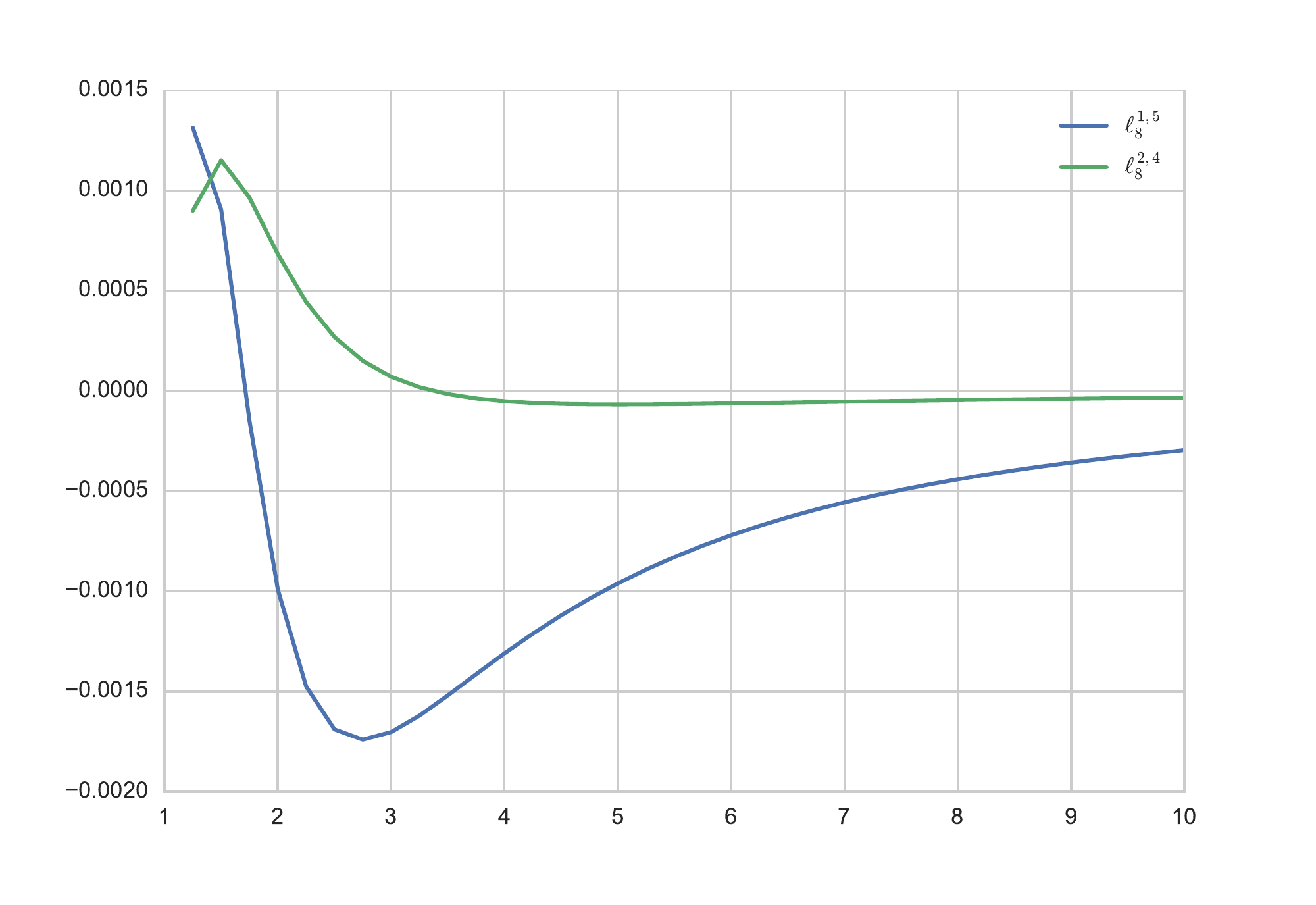}%
\label{fig61}} \vspace{-1ex}
\subfigure[Unbalanced $\ell$-graphs of order $9$]{\includegraphics[width=3.2in]{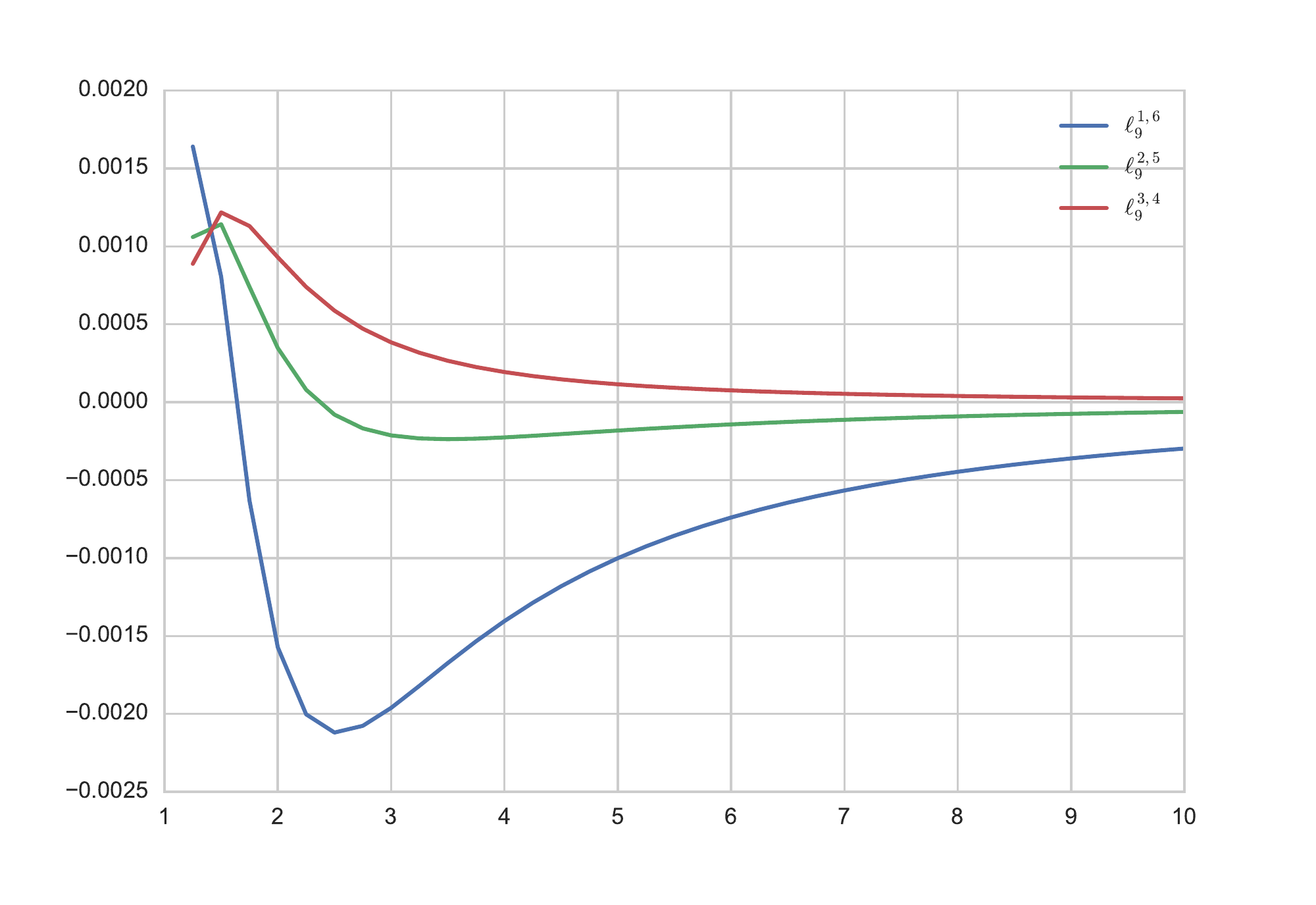}%
\label{fig62}}  \vspace{-1ex}
\subfigure[Unbalanced $\ell$-graphs of order $10$]{\includegraphics[width=3.2in]{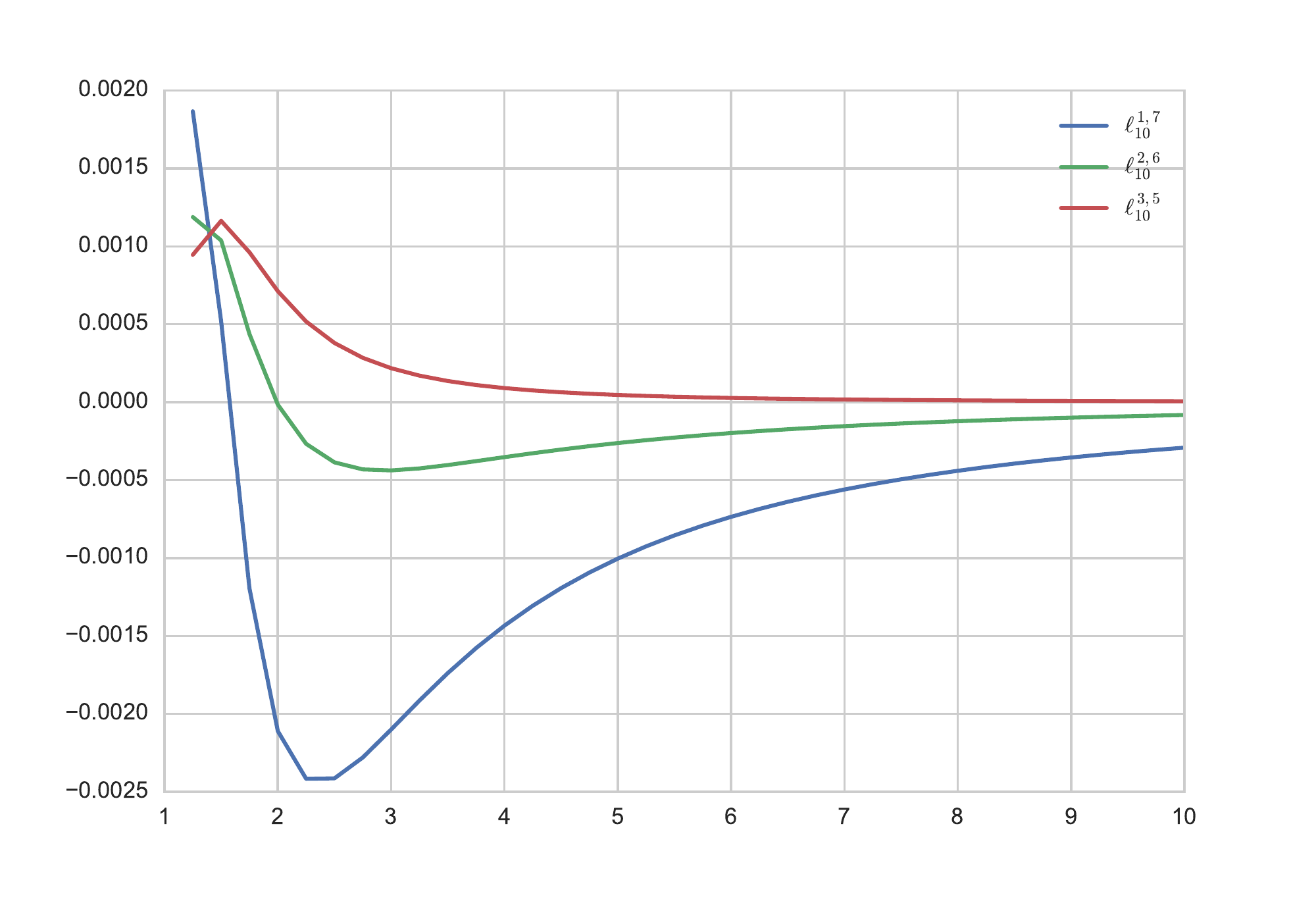}%
\label{fig63}}
 \caption{\bf The differences $\Delta(r) = \Phi(r) - \Phi_0(r)$ for the unbalanced graphs of order $8$, $9$, and $10$.} 
  \label{fig6}
\end{figure}

To investigate the structural reasons of the suppression of selection in these graphs, this experiment has been completed 
 by altering the balance in the connections of the two extra nodes with the central complete graph in order $6$ and considering two variants (\emph{a fortiori} unbalanced) of order $7$ (see Figure~\ref{unbalanced}). As showed in Figure~\ref{fig5}, the graphs $\ell_6^{1,3}$ and $\ell_7^{1,4}$ become amplifiers of selection from relatively small values of the fitness, while the graph $\ell_7^{2,3}$ is a suppressor of selection for high fitness values. We discover a similar behaviour for larger orders (see Figure~\ref{fig6}).


\section{Discussion}
Motivated by interest in the robustness of networks against invasion,
we worked with the \emph{FinisTerrae2} supercomputer (using 1024 cores of Haskell 2680v3 CPUs for almost 3 days) installed at CESGA to compute the fixation probability of all undirected graphs of order $10$ or less
for fitness values $r$ varying from $0.25$ to $10$ with step size of $0.25$ \cite{AGSL2}. Exploring these data, it would be possible to shed some light on the influence of the structural properties of graphs upon increasing or decreasing the fixation probability of new invaders occupying the nodes of a network. In this paper, we proved that there are graph structures acting as suppressors of selection according to the terminology introduced in \cite{LHN,N}. This means that, for every fitness value $r >1$, the average fixation probability $\Phi(r)$ of an advantageous invader individual placed at a random node is strictly less than that of this individual placed in a well-mixed population. For neutral drift $r =1$, both probabilities $\Phi(1)$ and $\Phi_0(1)$ are obviously equal, whereas the average fixation probability $\Phi(r)$ becomes strictly greater than $\Phi_0(r)$ for a disadvantageous invader with fitness $r <1$. We proposed a novel method to compute the fixation probability of graphs having low order and a big group of symmetries, and we used computer aided techniques to find an exact analytical expression of the fixation probability for three examples of size $6$, $8$ and $10$. A SageMath program \cite{sage} to compute the fixation probability of these graphs is included in the Supporting Information of the paper. Monte Carlo simulation was also used to see with high precision that other graphs in this family are suppressors of selection for some fitness values (varying from $1$ to $4$ with step size of $0.25$). Memory requirements make unfeasible to apply the same method for large orders, but it could be useful to study transitions between both regimes, suppression and amplification, in low order. On the other hand, although we are only concerned here with the evolutionary dynamics of graphs under birth-death updating, similarly to the work by Kaveh et al. \cite{KKK} and Hindersin et al. \cite{HindersinTraulsen}, it could be also interesting to study the properties of the $\ell$-family under death-birth updating.  We also showed that the mechanism that activates the suppression of selection is quite subtle, since a certain imbalance in the number of nodes of the central complete graph which are connected with each additional node transforms our models into amplifiers from certain fitness values.
 Finally, if the spreading of favorable innovations can be enhanced by those network structures amplifying the advantage of mutant or invader individuals \cite{TanLu}, as counterpart, the discovery of these examples is a first step towards finding structural properties that increase the robustness of a complex network against invasion \cite{SciRep}. This is a particularly interesting property for biological networks like brain networks or PPI interactomes, as well as for technological networks like electrical power grids or backbone networks,  where high fitness values are possible. In fact, these kind of models have had impact not only in evolutionary and invasion dynamics, but also in tumor growth \cite{N,HWDT,KSN} and economics and management \cite{SSL}. 

\section{Methods} 

\subsection{Mathematical model}
Let $G$ be a connected undirected graph with node set 
$V=\{1,\dots,N\}$. Denote by $d_i$ the degree of the node $i$.
The \emph{Moran process} on $G$ is a Markov chain $X_n$ whose 
states are the sets of nodes $S$ inhabited by mutant or invader individuals at each time step $n$. The transition 
probabilities are obtained from a stochastic matrix 
$W = (w_{ij})$ where $w_{ij} = 1/d_i$ if $i \sim j$ and $w_{ij} = 0$ otherwise. More precisely, 
the transition probability between 
$S$ and $S'$ is given by
$$
P_{S,S'} =  
\left\{
\begin{array}{ll}
\frac{r \sum_{i \in S} w_{ij}}{w_G} & \text{\footnotesize if $S' \rs S = \{j\}$,}\\
\frac{\sum_{i \in V \rs S} w_{ij}}{w_G} & \text{\footnotesize  if $S \rs S' = \{j\}$,} \\
\frac{r \sum_{i,j \in S} w_{ij} + \sum_{i,j \in V \rs S} w_{ij}}{w_G} & \text{\footnotesize  if $S=S'$,} \\
{\scriptstyle 0} & \text{\footnotesize  otherwise,}
\end{array} \right.
$$
where $r>0$ is the fitness and 
$$
w_G  = r \sum_{i\in S} \sum_{j\in V} w_{ij} + \sum_{i\in V\rs S} \sum_{j\in V} w_{ij}  =  r \num{S} + N - \num{S} 
$$
 is the total reproductive weight of invaders and residents. 
The \emph{fixation probability} of each subset $S \subset V$ inhabited by invaders 
$\Phi_S(r) =  \mathbb{P} \, [ \, \exists n \geq 0 : X_n = V \, | \, X_0 = S \,]$ gives a solution of the system of $2^N$ linear equations 
\begin{equation}  \label{eq:fixation}
\Phi_S(r) = \sum_{S'} P_{S,S'} \Phi_{S'}(r).
\end{equation}
Since $G$ is undirected, the only recurrent states are $S = \emptyset$ and $S = V$. Then
Eq~\eqref{eq:fixation} has a unique solution  \cite{KT:ISM}.
The \emph{(average)
fixation probability} is given by 
\begin{equation}  \label{eq:average}
\Phi(r) = \frac1N \sum_{i=1}^N \Phi_{\{i\}}(r).
\end{equation} It is a rational function depending on the fitness $r \in (0,+\infty)$. Notice that $\Phi(r)$ may be calculated using  the embedded Markov chain instead of the standard Markov chain above described, both associated to the process, making the total reproductive weight disappear from the computations \cite{AGSL}.

\subsection{Computation method}

As we remarked above, the average fixation probability is a rational function $\Phi(r) = \Phi'(r)/\Phi''(r)$ where the numerator $\Phi'(r)=\sum_{i} a_i r^i$ and the denominator $\Phi''(r)=\sum_{i} b_i r^i$ are 
polynomials with rational coefficients of degree less than or equal to $2^N-2$. Using the symmetries of each $\ell$-graph,  we can reduce the space of states $\mathcal{P}(V)$ to the set of $4$-uplas
$$
(e,k,k',e') \in \{0,1\}\times \{0,1,\dots,n\} \times  \{0,1,\dots,n\} \times  \{0,1\}
$$
ordered lexicographically (from halves to extra vertices) by $ k \geq k'$ or $e \geq e'$, or equivalenty the system of linear equations Eq~\eqref{eq:fixation} to a new system with at most 
\begin{eqnarray*}
\frac{(2(n+1))^2 + 2(n+1)}{2}  =  \frac{N^2 + N}{2} =  \frac{N(N+1)}{2} 
\end{eqnarray*}
 linear equations. For $\ell_6$, $\ell_8$ and $\ell_{10}$, we have $21$, $36$ and $55$ reduced states respectively. 
Therefore, we can lower the former bound of the degree of $\Phi'(r)$ and $\Phi''(r)$ to a quantity 
$$
d = \frac{N(N+1)}{2} - 2
$$
proving Eq~\eqref{eq:d}. Hence, we should only compute the $2(d+1)$ coefficients involved in $\Phi(r)$. Actually, since $\Phi(r)$ converges to $1$ as $r \to+\infty$, we can assume that $a_d=b_d=1$. Thus, we can replace Eq~\eqref{eq:fixation} with the system of $2d$ linear equations
\begin{equation} \label{eq:reduced}
  \sum_{i=0}^d a_i r^i = \Phi(r)(\sum_{i=0}^d b_i r^i),
\end{equation}
which arise from evaluating the rational function $\Phi(r)$ for fitness values
$r \in \{ 1, \ldots, d+1, 1/2, \ldots, 1/d\}$. 
This choice is due to those are the least complex rational numbers, 
which can be described with only few bits, and the length in bits of the solution of Eq~\eqref{eq:reduced} grows exponentially depending on the coefficients \cite{FangHavas}. Finally, we wrote a SageMath program \cite{sage}
\begin{itemize}

\item to compute the exact fixation probability $\Phi(r)$ of the graphs $\ell_6$, $\ell_8$ and $\ell_{10}$ for these fitness values, 

\item to solve the reduced linear system \eqref{eq:reduced}.

\end{itemize} 
This program is included in the Supporting Information of the paper.
Once the fixation function $\Phi$ of the graphs $\ell_6$, $\ell_8$ and $\ell_{10}$ has been calculated solving this system, the sign of the numerator $\Delta'$ and the denominator $\Delta''$ of the rational function 
$\Delta(r)= \Phi(r) - \Phi_0(r)$ is analyzed in order to prove that $\Delta(r) < 0$ for all $r >  1$. The exact values of $\Phi$ and $\Delta$ are also given in the Supporting Information.

\section*{Acknowledgements} 
\noindent
This work was supported by the 
Spanish Ministry of Economy and Competitiveness (Grant MTM2013-46337-C2-2-P), the
Government of Galicia (Grant GPC2015/006) and the European Regional Development.
Third author was also supported by the Government of Arag\'on and the European Regional Development Fund (Grant E15 Geometr\'{\i}a) and the Defense University Center of Zaragoza (Grant CUD 2015-10). The authors thank the supercomputer facilities provided by CESGA.

\newpage

\newgeometry{right=1.1in,left=1.1in, top=1in,footskip=30pt}



\section*{Supplementary material}

# Supporting Information

As supplementary information, we include here the text of the SageMath program which we ran
- to compute the exact fixation probability $\Phi(r)$ for the graphs $\ell_6$, $\ell_8$ and $\ell_{10}$ when the fitness values $r \in \{1, \ldots, d+1, 1/2, \ldots, 1/d\}$,
- to solve the linear system

$$\sum_{i=0}^{d} a_i r^i = \Phi(r)\bigg(\sum_{i=0}^{d} b_i r^i\bigg),$$

where $\Phi'(r) = \sum_{i=0}^{d} a_i r^i$ and $\Phi''(r) = \sum_{i=0}^{d} b_i r^i$ are the numerator and the denominator of the rational function $\Phi(r)$. The symmetries of the graph allow to reduce the degree which is now bounded by $d = \frac{N(N+1)}{2} - 2 \ll 2^N - 2$. Next, we give the exact values of $\Phi = \Phi'/\Phi''$ and $\Delta = \Delta'/\Delta''$ for the graphs $\ell_6$, $\ell_8$ and $\ell_{10}$.

## Text of the SageMath program

```
#
# The l-family checker
#
# The graph l_{2n+2} is a complete graph K_{2n} with two more nodes
# each of them connected with half the nodes of the complete core.
# These new nodes are also joint by an edge.
#
# The nodes are writen as
#
#       (e,k,k',e')
#
# where e,e'\in\{0,1\} represents if the external nodes are mutants or not,
# and k,k'\in{0,1,...,n} the number of mutants of each halves of the complete
# core. Using the symmetries of the graph, it is possible to reduce the nodes
# to those with (lexicographically)
#
#       k >= k'                 e >= e'
#

# The parameter n is:
n = 2 # change it to 3 and 4 to reproduce the results of the paper


#
# Reduces the state s to a canonical form
#
def reduce_state(s):
    if s[1] < s[2]:
        return (s[3],s[2],s[1],s[0])
    if (s[1] == s[2]) and (s[3] > s[0]):
        return (s[3],s[2],s[1],s[0])
    return s
```

```python
#
# Computes the possible (reduced) states for a given n, it returns the list of
# states and the position of the states in the list
#
def ComputeStates(n):
    # Compute the states
    states = []
    for k in [0..n]:
        for kp in [0..k]:
            if k != kp:
                states.extend([ (0,k,kp,0), (1,k,kp,0), (0,k,kp,1), (1,k,kp,1) ])
            else:
                states.extend([ (1,k,kp,1), (1,k,kp,0), (0,k,kp,0) ])

    istates = { v:k for k,v in enumerate(states)}

    s0 = states[0]
    sn = states[-1]
    states[0] = (0,0,0,0)
    states[-1] = (1,n,n,1)
    states[istates[(0,0,0,0)]] = s0
    states[istates[(1,n,n,1)]] = sn
    istates[s0] = istates[(0,0,0,0)]
    istates[sn] = istates[(1,n,n,1)]
    istates[(0,0,0,0)] = 0
    istates[(1,n,n,1)] = len(states)-1
    return states, istates



#
# Computes the matrix P and the vector b for the list
#
def ComputePb(r,n,states,istates):

    # Compute matrix
    P = matrix(PP,[[0]*len(states)]*len(states) )

    for s in states:
        m = sum(s)
        P[istates[s],istates[s]] = r * m + 2 * n + 2 - m


    for s in states:
        # Leftmost
        if s[0] == 1:
            P[istates[s],istates[(1,s[1],s[2],1)]] -= r/(n+1)
```

```python
        if s[1] == n:
            P[istates[s],istates[s]] -= r*n/(n+1)
        else:
            P[istates[s],istates[(s[0],s[1]+1,s[2],s[3])]] -= r*(n-s[1])/(n+1)
            P[istates[s],istates[s]] -= r*s[1]/(n+1)
    else:
        P[istates[s],istates[(0,s[1],s[2],0)]] -= 1/(n+1)

        if s[1] == 0:
            P[istates[s],istates[s]] -= n/(n+1)
        else:
            P[istates[s],istates[reduce_state((s[0],s[1]-1,s[2],s[3]))]] -= s[1]/(n+1)
            P[istates[s],istates[s]] -= (n-s[1])/(n+1)


    # Rightmost
    if s[3] == 1:
        P[istates[s],istates[(1,s[1],s[2],1)]] -= r/(n+1)
        if s[2] == n:
            P[istates[s],istates[s]] -= r*n/(n+1)
        else:
            P[istates[s],istates[reduce_state((s[0],s[1],s[2]+1,s[3]))]] -= r*(n-s[2])/(n+1)
            P[istates[s],istates[s]] -= r*s[2]/(n+1)
    else:
        P[istates[s], istates[(0,s[1],s[2],0)]] -= 1/(n+1)

        if s[2] == 0:
            P[istates[s],istates[s]] -= n/(n+1)
        else:
            P[istates[s],istates[(s[0],s[1],s[2]-1,s[3])]] -= s[2]/(n+1)
            P[istates[s],istates[s]] -= (n-s[2])/(n+1)


    # Core left
    if s[1] == n:
        P[istates[s],istates[(1,n,s[2],s[3])]] -= r/2
        P[istates[s],istates[s]] -= r*(n-1)/2
        if s[2] == n:
            P[istates[s],istates[s]] -= r*n/2
        else:
            P[istates[s],istates[reduce_state((s[0],n,s[2]+1,s[3]))]] -= r*(n-s[2])/2
            P[istates[s],istates[s]] -= r*s[2]/2
    elif s[1] == 0:
        P[istates[s],istates[(s[3],0,0,0)]] -= 1/2
        P[istates[s],istates[s]] -= (n-1)/2 + n/2
    else:
        P[istates[s],istates[(1,s[1],s[2],s[3])]] -= s[1]*r/(2*n)
        P[istates[s],istates[reduce_state((0,s[1],s[2],s[3]))]] -= (n-s[1])/(2*n)
        P[istates[s],istates[s]] -= s[1]*r*(s[1]-1)/(2*n) + (n-s[1])*(n-s[1]-1)/(2*n)
```

```python
                P[istates[s],istates[(s[0],s[1]+1,s[2],s[3])]] -= s[1]*r*(n-s[1])/(2*n)
                P[istates[s],istates[reduce_state((s[0],s[1]-1,s[2],s[3]))]] -= (n-s[1])*s[1]/(2*n)
                P[istates[s],istates[s]] -= r*s[1]*s[2]/(2*n) + (n-s[1])*(n-s[2])/(2*n)
                P[istates[s],istates[reduce_state((s[0],s[1],s[2]+1,s[3]))]] -= r*s[1]*(n-s[2])/(2*n)

                if s[2] != 0:
                    P[istates[s],istates[(s[0],s[1],s[2]-1,s[3])]] -= (n-s[1])*s[2]/(2*n)

            # Core right
            if s[2] == n:
                P[istates[s],istates[(1,n,n,s[0])]] -= r/2
                P[istates[s],istates[s]] -= r*(2*n-1)/2
            elif s[2] == 0:
                P[istates[s],istates[(s[0],s[1],0,0)]] -= 1/2

                if s[1] == 0:
                    P[istates[s],istates[s]] -= (2*n-1)/2
                else:
                    P[istates[s],istates[s]] -= (n-1)/2 + (n-s[1])/2
                    P[istates[s],istates[reduce_state((s[0],s[1]-1,0,s[3]))]] -= s[1]/2
            else:
                P[istates[s],istates[reduce_state((s[0],s[1],s[2],1))]] -= s[2]*r/(2*n)
                P[istates[s],istates[(s[0],s[1],s[2],0)]] -= (n-s[2])/(2*n)
                P[istates[s],istates[s]] -= s[2]*r*(s[2]-1)/(2*n) + (n-s[2])*(n-s[2]-1)/(2*n)
                P[istates[s],istates[s]] -= r*s[2]*s[1]/(2*n) + (n-s[2])*(n-s[1])/(2*n)
                P[istates[s],istates[reduce_state((s[0],s[1],s[2]+1,s[3]))]] -= s[2]*r*(n-s[2])/(2*n)
                P[istates[s],istates[(s[0],s[1],s[2]-1,s[3])]] -= (n-s[2])*s[2]/(2*n)
                if s[1] != n:
                    P[istates[s],istates[(s[0],s[1]+1,s[2],s[3])]] -= r*s[2]*(n-s[1])/(2*n)
                    P[istates[s],istates[reduce_state((s[0],s[1]-1,s[2],s[3]))]] -= (n-s[2])*s[1]/(2*n)

    b=-P[1:len(states)-1,len(states)-1]
    P=P[1:len(states)-1,1:len(states)-1]
    return P,b


#
# Computes the average fixation probability function
#
def Phi(P,b,n,istates):
    sol = P.solve_right(b)
    return sol[istates[(1,0,0,0)]-1,0] / (n + 1) + sol[istates[(0,1,0,0)]-1,0] * n / (n + 1)


#
# "Entry point"
#
```

```python
# Computes the states for the given size
states, istates = ComputeStates(n)

# A first estimate of the degree of the rational function Phi
num_fac = len(states) - 2
fits=[1..num_fac+1]
fits.extend([ 1/f for f in fits[1:-1] ])

# Compute the fixation probabilities for the selected fitnesses
fps = {
    f : Phi(*ComputePb(f, n, states, istates), n=n, istates=istates)
    for f in fits
}

# Now we are ready to compute the coefficients of the function.
# Construct the matrix of the linear system describing the function \Phi
X = matrix(PP, [[1]*len(fits)]*len(fits) )
X[:,num_fac-2] = [ [r] for r in fits ]
X[:,-1] = [ [ -fps[r] ] for r in fits ]
X[:,-2] = [ [ -r*fps[r] ] for r in fits ]

y = vector(PP, [(fps[r]-1)*r**num_fac for r in fits ] )

for i in xrange(num_fac-3,-1,-1):
    X[:,i] = X[:,i+1].elementwise_product( X[:,num_fac-2] )
    X[:,num_fac + i] = X[:,num_fac+i+1].elementwise_product( X[:,num_fac-2] )

# Solve it
sol = X.solve_right(y)

# Create the field of fuctions. Then x is regarded as the variable of the
# field P(x). You can bypass this step to simplify later the expressions
# (that is regarding x as a symbolic variable with no special meaning)
F.<x> = Frac(PP['x'])

# Construct the function
FB = (sum( sol[num_fac-1-i]* x**i for i in [0..num_fac-1] ) + x**num_fac) / \
                    (sum( sol[-(i+1)]* x**i for i in [0..num_fac-1] ) + x**num_fac)
pretty_print(FB)

# and finally the Delta
Delta = FB - x**(2*n+1)/sum( x**i for i in range(2*n+2) )
pretty_print(Delta)

# take a look to the numerator and denominator
pretty_print(Delta.denominator())
pretty_print(Delta.numerator()/(x-1)) # up to (x-1)
```

# Fixations functions for $\ell_6$, $\ell_8$ and $\ell_{10}$

**Graph $\ell_6$**

$\Phi'(r) = r^{17} + \frac{20010803101}{2356885440}r^{16} + \frac{12872954057}{368903808}r^{15} + \frac{146529682470259}{1590897672000}r^{14} + \frac{20006411662050293}{114544632384000}r^{13} + \frac{57705463723529081}{229089264768000}r^{12} + \frac{9771555843017869}{34363389715200}r^{11} + \frac{1025122250624863}{4042751731200}r^{10} + \frac{60836752923313811}{343633897152000}r^9 + \frac{9233243223215201}{98181113472000}r^8 + \frac{1542392404026107}{42954237144000}r^7 + \frac{3153322564729}{357951976200}r^6 + \frac{496442293}{474304896}r^5$

$\Phi''(r) = r^{17} + \frac{22367688541}{2356885440}r^{16} + \frac{1883112090533}{42423937920}r^{15} + \frac{579196292648299}{4242393792000}r^{14} + \frac{23779172425823219}{76363088256000}r^{13} + \frac{17230349794290811}{30545235302400}r^{12} + \frac{55592085032472109}{65454075648000}r^{11} + \frac{71798191286820983}{65454075648000}r^{10} + \frac{37885437915832519}{30545235302400}r^9 + \frac{37885437915832519}{30545235302400}r^8 + \frac{71798191286820983}{65454075648000}r^7 + \frac{55592085032472109}{65454075648000}r^6 + \frac{17230349794290811}{30545235302400}r^5 + \frac{23779172425823219}{76363088256000}r^4 + \frac{579196292648299}{4242393792000}r^3 + \frac{1883112090533}{42423937920}r^2 + \frac{22367688541}{2356885440}r + 1$

$\Delta'(r) = r^5(r-1)(-\frac{1}{408}r^{13} - \frac{17959403797}{509087255040}r^{12} - \frac{2269495014623}{9163570590720}r^{11} - \frac{165933379973713}{152726176512000}r^{10} - \frac{4365681088426001}{1374535588608000}r^9 - \frac{4487811876686051}{687267794304000}r^8 - \frac{6774104367192077}{687267794304000}r^7 - \frac{673669214287879}{59762416896000}r^6 - \frac{13704670736178649}{1374535588608000}r^5 - \frac{2361661570426121}{343633897152000}r^4 - \frac{2506212053017441}{687267794304000}r^3 - \frac{122207986609099}{85908474288000}r^2 - \frac{299174081287}{818175945600}r^1 - \frac{22137397}{474304896})$

$\Delta''(r) = r^{21} + \frac{22367688541}{2356885440}r^{20} + \frac{1925536028453}{42423937920}r^{19} + \frac{619458132022099}{4242393792000}r^{18} + \frac{27245137277038619}{76363088256000}r^{17} + \frac{15493177389178517}{21818025216000}r^{16} + \frac{552157240360000477}{458178529536000}r^{15} + \frac{411797892764062669}{229089264768000}r^{14} + \frac{550050599259865931}{229089264768000}r^{13} + \frac{1329324154659596831}{458178529536000}r^{12} + \frac{540745741841168127}{16969575168000}r^{11} + \frac{540745741841168127}{16969575168000}r^{10} + \frac{1329324154659596831}{458178529536000}r^9 + \frac{550050599259865931}{229089264768000}r^8 + \frac{411797892764062669}{229089264768000}r^7 + \frac{552157240360000477}{458178529536000}r^6 + \frac{15493177389178517}{21818025216000}r^5 + \frac{27245137277038619}{76363088256000}r^4 + \frac{619458132022099}{4242393792000}r^3 + \frac{1925536028453}{42423937920}r^2 + \frac{22367688541}{2356885440}r + 1$

**Graph $\ell_8$**

$\Phi'(r) = r^{31} + \frac{134280899640541}{8309364148800}r^{30} + \frac{51436353602527862867}{399248328621542400}r^{29} + \frac{2613952394818525455380878881}{3873507284286204364800000}r^{28} + \frac{583807879596527795907799931453}{22369504566752830206720000}r^{27} + \frac{27324428550951402003280646028 7}{344146224103889695488000000}r^{26} + \frac{10954679493630587091709024774 49001}{5541864357182636644761600000}r^{25} + \frac{103820319576966457199577759964 9}{25100123467406200012800000}r^{24} + \frac{9219730675269962838432389793116 41}{12437849417025284053405340160000}r^{23} + \frac{11740712284825831069430912101427861}{10180610078379954725191680 0000}r^{22} + \frac{1032937952392948232127781620508111}{2080310449462551016365104268144504161}r^{21} + \frac{2080310449462551016365104268144504161}{10995058884650351103207014400000}r^{20} + \frac{753897025982731385467385013243433}{37398159471599833684377 60000}r^{19} + \frac{2463546677977027065113743507694358 1}{12935363393706295415537664 0000}r^{18} + \frac{12515846318543399852954221 13577}{7853613489035965073719296000}r^{17} + \frac{347790118541169116877441122 4802753}{29556609904974062105395200 000}r^{16} + \frac{5510437078852008671225781 612417}{72335913714804941468467200 000}r^{15} + \frac{47090311290917554231879957186 9746773}{10995058884650351103207014400000}r^{14} + \frac{596923826377497575290462451325719}{7066876349108602065069037131 219649}r^{13} + \frac{7066876349108602065069037131 219649}{84577376035719315631308800 00}r^{12} + \frac{15230658652580143385602731564041827}{54975299442325717551603507200 000}r^{11} + \frac{14193002742417197051814844712131}{19634033722589912684298240 0000}r^{10} + \frac{911745030267142017364679213416 5773}{65446779075299709470805780 000}r^9 + \frac{624915283883754887942650 1}{35414923742045297049600 0}r^8 + \frac{11363885569460556 03}{10303182674104320}r^7$

$\Phi''(r) = r^{31} + \frac{142590263789341}{8309364148800}r^{30} + \frac{685751104869983351}{4697039160253440}r^{29} + \frac{3179702723486717658719761481}{3873507284286204364800000}r^{28} + \frac{6140454463158671891965 42724177}{178956036534022641653 760000}r^{27} + \frac{7828830358350018509 13227050807}{6882924482077793909 7600000}r^{26} + \frac{2027468204090641240495 620587659}{6507492237600823332 8640000}r^{25} + \frac{6233679485365536931871 855089022867}{8589889753633086 79938040 80000}r^{24} + \frac{54059973838852759542443 81013403073}{368138132985608628 3059200000}r^{23} + \frac{541223350230039703505 39078924384011}{2061573540871940 83185131520000}r^{22} + \frac{17335457980981817006 01364139420724667}{412314 70817438816637026 30400000}r^{21} + \frac{16776173149945946521412776 95694600403}{274876472116258777580175 3600000}r^{20} + \frac{1337864832697552665481 052160}{16492588326975526 65481052160000}r^{19} + \frac{1066417788134171727 1399745414940 4519}{1070947293 95944978 277990400000}r^{18} + \frac{937459 39780908581 52054 74615900095 3163}{824629 41634 87763327 405 260800000}r^{17} + \frac{3969 6361687927740 61518 6917132468 293}{3272338953 76498 544 73830400000}r^{16} + \frac{3969636 1687927 740615 18691713 2468293}{327233895 37649854473 830400000}r^{15} + \frac{93745939 78090858 152 0547 46159000 953163}{82462941634877633 27405260 800000}r^{14} + \frac{10664 17788 134171 727139 945414 9404519}{107094 729 3959 44978277 990 400000}r^{13} + \frac{13378 64897 7067 1953923 6595 66158 9955 8203}{16 49258 83269 7 552665 481052 160000}r^{12} + \frac{1677 61731 49945 94652141 2776956946 0040 3}{27487 64721 162587 7758017536 00000}r^{11} + \frac{1733 5457980981 81700601 36413 9420 724 667}{41 23147 08174388166 37026 30 400 000}r^{10} + \frac{54122 3350 23003970350 53907 89243 84011}{2061 573540 87 194083 185131 520000}r^9 + \frac{5405 9973 838852 759 5424 438101340 3073}{3681381329856 0 862830592 00000}r^8 + \frac{62336794853655 36931871 8550 890228 67}{858988 9753633086 799380480 00}r^7 + \frac{20274 682040 906412 40 495 620 587659}{6507 49223 76008 2333 28640 000}r^6 + \frac{782883035 8350018509 1322705 0807}{68 82924482 0777793909 7600 000}r^5 +$

17

$$\frac{6140454463156862381955427241\overline{77}}{17895603653402264165376000\overline{0}}r^4 + \frac{317970272348671658719761481}{387350728428620436480000}r^3 + \frac{685751104869983351}{4697039160253440}r^2 + \frac{142590263789341}{8309364148800}r + 1$$

$$\Delta'(r) = r^7(r-1)(-\tfrac{1}{304}r^{27} - \tfrac{25197109151333}{398849479142400}r^{26} - \tfrac{79418092024874684869}{130663089367050240000}r^{25} - \tfrac{151439378562924022622196129}{387350728428620436480000}r^{24} - \tfrac{23669739102105610530381864\overline{71}}{124943850961935807990988800}r^{23} - \tfrac{8455314295207479062300968873111}{114531863381774490658406400000}r^{22} - \tfrac{9915244787998601128380203282098\overline{49}}{41231470817438816637026304000\overline{00}}r^{21} - \tfrac{55211648444687591594107305858690\overline{37}}{82462941634877633274052608000\overline{00}}r^{20} - \tfrac{9853645855648051908124376208625\overline{39}}{61083660470279728351150080000\overline{0}}r^{19} - \tfrac{111731529120083126099350945280644\overline{519}}{329851766539510533096210432000\overline{00}}r^{18} - \tfrac{2054609370880573116532051768359613\overline{21}}{329851766539510533096210432000\overline{00}}r^{17} - \tfrac{11073449367030373276155500111578751\overline{3}}{1099505888465035110320701440000\overline{0}}r^{16} - \tfrac{15803272413831999063713425526311023\overline{1}}{1099505888465035110320701440000\overline{0}}r^{15} - \tfrac{19980272099587550846028888561940723\overline{7}}{1099505888465035110320701440000\overline{0}}r^{14} - \tfrac{22453860958317641258746883635302901\overline{1}}{1099505888465035110320701440000\overline{0}}r^{13} - \tfrac{24094663993232446388630691852040643}{1178042023553947610578944000\overline{00}}r^{12} - \tfrac{15076786364712664621750564243779907\overline{7}}{82462941634877633274052608000\overline{00}}r^{11} - \tfrac{5345105200264500037417701584886285\overline{1}}{366501962821678370106900480000\overline{0}}r^{10} - \tfrac{3415779738547508098280948450069283\overline{7}}{329851766539510533096210432000\overline{00}}r^9 - \tfrac{6134982182129497498119956432252281}{942433618684315808463155200\overline{00}}r^8 - \tfrac{56390563795674089013546736289013\overline{77}}{1570722697807193014743859200000}r^7 - \tfrac{18840773531540731523638914353570237}{1099505888465035110320701440000}r^6 - \tfrac{218101089181656220622956529171233}{31414453956143860294877184000\overline{0}}r^5 - \tfrac{851361180489465323100617479312\overline{73}}{3665019628216783701069004800\overline{00}}r^4 - \tfrac{1329949113042587237018515742\overline{51}}{21575861233615288664064000\overline{0}}r^3 - \tfrac{796217688916573522810090562\overline{53}}{654467790752997089476608000\overline{0}}r^2 - \tfrac{56249336779522894155418\overline{1}}{35414923742045297049600\overline{0}}r - \tfrac{1060703020501283}{10303182674104320})$$

$$\Delta''(r) = r^{37} + \tfrac{142590263789341}{8309364148800}r^{36} + \tfrac{690448144030236791}{4697039160253440}r^{35} + \tfrac{324617284268784317447995081}{387350728428620436480000}r^{34} + \tfrac{640351350752994828494392138177}{17895603653402264165376000\overline{0}}r^{33} + \tfrac{10927345392515915835176820297601}{89478018267011320826880000\overline{0}}r^{32} + \tfrac{30695747713591398715344932119\overline{7}}{883733513748259958784000\overline{0}}r^{31} + \tfrac{34845476760990740918893922667907}{410999509743209894707200000}r^{30} + \tfrac{46792604023787574552501267885526031}{257696692608992603981414400\overline{00}}r^{29} + \tfrac{51140920333603397111269496525056973}{147255252919424345132236800\overline{00}}r^{28} + \tfrac{82720849919587634913418754590699372\overline{1}}{82462941634877633274052608000\overline{00}}r^{27} + \tfrac{7889974227256354217659306239108441937}{82462941634877633274052608000\overline{00}}r^{26} + \tfrac{11624280269542628969132052974659677181}{82462941634877633274052608000\overline{00}}r^{25} + \tfrac{19056661363252567398670117026320839\overline{1}}{98170186129495634214912000\overline{00}}r^{24} + \tfrac{7856800559398151917875452307875940\overline{1}}{31235962740483951997747200\overline{00}}r^{23} + \tfrac{10588602274956189876854293054560549\overline{77}}{34359559014532347197521920000\overline{0}}r^{22} + \tfrac{738362330204465247111688896579305083\overline{7}}{2061573540871940831851315200\overline{00}}r^{21} + \tfrac{17169655808981871074991308596300480\overline{09}}{4340154822888296488108032000\overline{00}}r^{20} + \tfrac{34278818580615368782242631553834901\overline{977}}{82462941634877633274052608000\overline{00}}r^{19} + \tfrac{34278818580615368782242631553834901\overline{977}}{82462941634877633274052608000\overline{00}}r^{18} + \tfrac{17169655808981871074991308596300480\overline{09}}{4340154822888296488108032000\overline{00}}r^{17} + \tfrac{7383623302044652471116889657930508\overline{37}}{206157354087194083185131520000\overline{0}}r^{16} + \tfrac{1058860227495618987685429305456054\overline{977}}{34359559014532347197521920000\overline{0}}r^{15} + \tfrac{7856800559398151917875452307875940\overline{1}}{31235962740483951997747200\overline{00}}r^{14} + \tfrac{19056661363252567398670117026320839\overline{1}}{98170186129495634214912000\overline{00}}r^{13} + \tfrac{11624280269542628969132052974659677181}{82462941634877633274052608000\overline{00}}r^{12} + \tfrac{7889974227256354217659306239108441937}{82462941634877633274052608000\overline{00}}r^{11} + \tfrac{82720849919587634913418754590699372\overline{1}}{82462941634877633274052608000\overline{00}}r^{10} + \tfrac{51140920333603397111269496525056973}{147255252919424345132236800\overline{00}}r^9 + \tfrac{46792604023787574552501267885526031}{257696692608992603981414400\overline{00}}r^8 + \tfrac{34845476760990740918893922667907}{410999509743209894707200000}r^7 + \tfrac{30695747713591398715344932119\overline{7}}{883733513748259958784000\overline{0}}r^6 + \tfrac{10927345392515915835176820297601}{89478018267011320826880000\overline{0}}r^5 + \tfrac{640351350752994828494392138177}{17895603653402264165376000\overline{0}}r^4 + \tfrac{324617284268784317447995081}{387350728428620436480000}r^3 + \tfrac{690448144030236791}{4697039160253440}r^2 + \tfrac{142590263789341}{8309364148800}r + 1$$

**Graph $\ell_{10}$**

$\Phi'(r) = r^{49} + \frac{639983374672251039920905452759 49}{245378435104996693552026824850 0}r^{48} + \frac{146779191511973282687623624712046560107 8713}{43471730639394897648113773063673327 25000}r^{47} + \frac{35863545786298851175250332766194774847587066497 001}{12401832675459773375542137748470045164516 250000}r^{46} + \frac{1010722502071856154011824311272938025729799016077 2275379}{5484834519048839373067265840638362174458956 72500000}r^{45} + \frac{173977361125932121188499854823535587602545984003692318 94071}{186745556243805721511575956002687093082769240 875000000}r^{44} + \frac{15848201759536847393147645085686097514735091342063338201 65933161}{4072981476967482288033181896925997680709028481279 687500000}r^{43} + \frac{517274463035147513860432834351215563141952618120529963719 9565311977}{3747142958810083704990527345171917866252306202777312 500000000}r^{42} + \frac{495212029559813543000300871191878853169935608075171197918 6627645861149}{116654476849271816394836680245746811467802058891725 2812500000000}r^{41} + \frac{2328196295056352042611226986974972135301795301844387586510 42497936165479}{20262186184080608407386744274028037588903678512723178 515625000000}r^{40} + \frac{14070516194229359931408566837646261771263445796257317605068 7488215725213}{50788121000127226460714996870926992834139411161707929 68750000000}r^{39} + \frac{70477610903605497182923534672076661678981050948653575414796 4424687649474571}{11732055951029389312425164277184135344686203978354531 75781250000000}r^{38} + \frac{56092389787989020774821289767346120935042736323208288149557 84214713994109652 13}{475148266016690267153219153225957481459791261123358536 19140625000000}r^{37} + \frac{50266749845106143928653677124964075300040173421165458420373 46221098565182809 09}{237574133008345133576609576612978740729895630561679268 09570312500000000}r^{36} + \frac{26429824015453355688749171370572651127697436882617726500323 253117106612145 203}{760237225626704427445150645161531970335666017797373657 9062500000000}r^{35} + \frac{20823387297836818281526008435918371919883630252711427765368 88747291708244698 9053}{39595868834724188929434929435496456788315938426946546 82617187500000}r^{34} + \frac{21824947647421174859349817853336560263525736968004742113593 6965629778268351 40353}{296967666260431416970761970766223425912369538202099085 1196289062500000}r^{33} + \frac{12073616682732932701560441487145827994410542880111328739604 0399570662572 5384629}{12690925908565445169690682511377069483434595649662354 0649410625000000}r^{32} + \frac{13578899405663158405855325658578217405476619450563723553598 49669605940440 768457227}{118787066504172566788304788306489370364947815280839634 04785156250000000}r^{31} + \frac{12639386650108450915282541313207950204672902015233103718834 84596474040356 624342127}{98989222086810472323587323588741141970789846067366361 70654298675000000 00}r^{30} + \frac{38937683392585997984347289370749507251595629992719108646252 0859044296962 887753}{29344631053402313929917190787176227856953511680049316 7114257812500000}r^{29} + \frac{12168244863533612374634189985671509035799081747091283158910 55585405587 207}{94817262535259073106884409567759714531407898531960 1165771484375000}r^{28} + \frac{34295065886761457556477474487433150078658942510448184698908 4698332117314061 325299}{29696766626043141697076197076622342591236953820209908 5119628906250000 00}r^{27} + \frac{13662088991753822285221782312481729552697653909684280517637 020148411519805 0916799}{1414131744097292461765533194124873456725569229533805167 23632812500000000}r^{26} + \frac{10610372870896170772509656575515409475380951963052421445259 36273517097828255 03313}{141413174409729246176553319412487345672556922953380516 72363281250000000 0}r^{25} + \frac{49604479358020455452567822996992838345023525231064 37714700470547}{918835601053315027756070454103414065322925551367 880832672119140625}r^{24} + \frac{77235948565418872482912042666997849846926353384680 3360587426280 9039}{21519396105828363548605939910595900428432575232036 165558378906250 00}r^{23} + \frac{65251636522798956760961292727355168062914739270411 9960328347885247667 6692854839209}{296967666260431416970761970766223425912369538202099 0851196289062500000}r^{22} + \frac{36630247400067144620094696169744275733530561389031 328559133377417470303 1631539261}{296967666260431416970761970766223425912369538202099 0851196289062500000}r^{21} + \frac{16454944186681652259954214500879127107116089999419 971419463709015853368 829896609}{26049795286002755874628243049668721571260485807201 67413330078125000 0}r^{20} + \frac{34833556722763089659481832034690135656178814634338 44672444674298813595 385516279}{1187870665041725667883047883064893703649478152808 396340478515625000 00}r^{19} + \frac{36366071775932414505549205751279497624573418272160 719024002888028025852 967255521}{296967666260431416970761970766223425912369538202099 0851196289062500 00}r^{18} + \frac{15660984538868686032512195177922507302933744244599 5634292556806549235 60486349}{343712576690314140012455984683128965176353632178 355422592163085937 50}r^{17} + \frac{98499985290369097258351936904704912306334259367334 39335488624857067513 903961}{659928147245403148823915490591607613138598973782 442411376953125000 0}r^{16} + \frac{76122446716433949660024959757715351941849586983844 817283774206365930539 589}{179572284964735550700385167507920438949278632321753 037109375000000 00}r^{15} + \frac{17803318924131766945932798215577703144061315400365 5112339468489230079 7027}{174046983888897533755757931584599810058531597481083 7128906250000000}r^{14} + \frac{66682395619551684635995287222011710927482980788072 849075839304 2227}{326940892061421120983860113805954372335523574306 09085119628906250 000}r^{13} + \frac{50241672766902471992514270225360386829136967407023 3164892842561}{155686139076867200468504816098073510587405884493 59375000000}r^{12} + \frac{20309000357168189131016179693765660499567247658979 879117}{534469444908586947308721574295373048214348828125 00}r^{11} + \frac{22310940281865417950066567363779288792016190600106 379}{75358358271022928497217850902161732770498046875 000}r^{10} + \frac{21609304615059677300836598774088547 87}{18869431352372327272248476992968 75000}r^9$

$$\Phi''(r) = r^{49} + \tfrac{6645212181827507092761081352449}{245378435104996693552026824500}r^{48} + \tfrac{3171065904103137883917151919165713711668601}{86943461278789795296227546127346654500000}r^{47} + \tfrac{16155146680896107863213416098535762926181898572279}{49607330701839093502168550993880180658065000000}r^{46} + \tfrac{206852964702209044167807729914130260919958800413049507}{9538842641824068474899592766327586390363403000000}r^{45} + \tfrac{4504356383135350054250392516243576690843449502201016816503}{3921656681119920151743095076056428954738154058375000000}r^{44} + \tfrac{6867963997466308046319209185518123663135600673726493180976212363}{1362597439567303165451100852789788315000838619191750000000}r^{43} + \tfrac{269061418914533471263783319921967679903930817801083532938975715087}{1427483031927650935234486607684540139524688077248500000000}r^{42} + \tfrac{108723943325927772842888786215567008404159907683523224614502048985526090}{177314804810893160920151753975351534310591295154224275000000000}r^{41} + \tfrac{38291101140138887359782676914615116290259958340221391656013420168256811577}{21721063589334412212718589861758056295304743365639247368750000000000}r^{40} + \tfrac{985149580035467745564381467626453493111554941047572568222165490724828462}{21721063589334412212718589861758056295304743365639247368750000000000}r^{39} + \tfrac{2673272931851582683179127499561657906563793539122235491025348682439847920827}{2534124085422348091483835483877173234452220059324578859687500000000}r^{38} + \tfrac{42518374152733113982590202752420032976346330396623572748207372340067857217467}{19005930640667610686128766129038299258391650444934314476562500000000}r^{37} + \tfrac{828072648510561263189682099297410898323565661136075420452698337442115035231749}{19005930640667610686128766129038299258391650444934314476562500000000}r^{36} + \tfrac{1490429730095424643406531637361546843361274895460296542726484508905081552517489}{19005930640667610686128766129038299258391650444934314765625000000000}r^{35} + \tfrac{4155048488190944638832776774327706161798547725172699805670965166909440401004850}{31676551067779351143547943548397165430652750741557235746093750000000000}r^{34} + \tfrac{2213045670764163900424814663060931502782381407062999513660324702584561292165160}{10798824227652051526209562096808518513588922982581485498046875000000000}r^{33} + \tfrac{1784422236438075188537138900658095059296529954702205069475487313232307453857597981}{5939353325208628339415239415324468518247390764041981702392578125000000}r^{32} + \tfrac{3287151019212848217807984745591777366372276860411583588494571887934733431111199393}{7919137766944837785886985887099291357663187685389308936523437500000000}r^{31} + \tfrac{12891830160861887042312930556116527397598369437831253112410533891470144462896563509}{2375741330083451335766095766129787407298956305616792680957031250000000}r^{30} + \tfrac{114336470684914447136873270162070667589536866081143319491584296884973658874004621}{1696958092916750954118639832949848148070683075440566200683593750000000}r^{29} + \tfrac{18942233618940241647091803441935082761492815955915919186658211519315537547323846207}{2375741330083451335766095766129787407298956305616792680957031250000000}r^{28} + \tfrac{2142809702988037321831111243535822967065689108206676097250195586450143213837889653}{80102201970881790445574619857861753005403804085441236475941159125379234358364643}r^{27} + \tfrac{819221148304638391643481298665443933551364243316135407226565250000000000}{604383109262508854121504773367312968100251010809618792543994505733615433493289742}r^{26} + \tfrac{5939353325208628339415239415324468518247390764041981702392578125000000}{6043831092625088541215047733673129681002510108096187925439945057336154334932897427}r^{25} + \tfrac{5939353325208628339415239415324468518247390764041981702392578125000000}{60438310926250885412150477336731296810025101080961879254399450573361543349328974}r^{24} + \tfrac{5939353325208628339415239415324468518247390764041981702392578125000000}{80102201970881790445574619857861753005403804085441236475941159125379234358364643}r^{23} + \tfrac{8192211483046383916434812986654439335513642433161354072265625000000}{21428097029880373218311112435358229670656891082066760972501955864501432138378896531}r^{22} + \tfrac{23757413300834513357660957661297874072989563056167926809570312500000}{18942233618940241647091803441935082761492815955915919186658211519315537547323846207}r^{21} + \tfrac{23757413300834513357660957661297874072989563056167926809570312500000}{114336470684914447136873270162070667589536866081143319491584296884973658874004621}r^{20} + \tfrac{16969580929167509541186398329949848148070683075440566200683593750000}{128918301608618870423129305561165273975983694378312531124105338914701444628965635}r^{19} + \tfrac{2375741330083451335766095766129787407298956305616792680957031250000}{3287151019212848217807984745591777366372276860411583588494571887934733431113199393}r^{18} + \tfrac{79191377669448377858869858870992913576631876853893089365234375000000}{1784422236438075188537138900658095059296529954702205069475487313232307453857597981}r^{17} + \tfrac{5939353325208628339415239415324468518247390764041981702392578125000}{2213045670764163900424814663060931502782381407062999513660324702584561292165160}r^{16} + \tfrac{1079882422765205152620968085185135889229825814854980468750000000000}{41550484881909446388327767743277061617985477251726998056709651669094404010048507}r^{15} + \tfrac{3167655106777935114354794354839716543065275074155723574609375000000}{1490429730095424643406531637361546843361274895460296542726484508905081552517489}r^{14} + \tfrac{19005930640667610686128766129038299258391650444934314476562500000000}{828072648510561263189682099297410898323565661136075420452698337442115035231749}r^{13} + \tfrac{19005930640667610686128766129038299258391650444934314476562500000000}{42518374152733113982590202752420032976346330396623572748207372340067857217467}r^{12} + \tfrac{19005930640667610686128766129038299258391650444934314476562500000000}{2673272931851582683179127499561657906563793539122235491025348682439847920827}r^{11} + \tfrac{2534124085422348091483835483877173234452220059324578859687500000000}{985149580035467745564381467626453493111554941047572568222165490724828462}r^{10} + \tfrac{21721063589334412212718589861758056295304743365639247368750000000}{38291101140138887359782676914615116290259958340221391656013420168256811577}r^{9} + \tfrac{21721063589334412212718589861758056295304743365639247368750000000}{108723943325927772842888786215567008404159907683523224614502048985526090}r^{8} + \tfrac{1773148048108931609201517539735351534310591295154224275000000000}{269061418914533471263783319921967679903930817801083532938975715087}r^{7} + \tfrac{142748303192765093523448660768454013952468807724690318097621236}{6867963997466308046319209185518123663135600673726493180976212363}r^{6} + \tfrac{1362597439567303165451100852789788315000838619191750000000}{4504356383135350054250392516243576690843449502201016816503}r^{5} + \tfrac{3921656681119920151743095076056428954738154058375000000}{206852964702209044167807729914130260919958800413049507}r^{4} + \tfrac{9538842641824068474899592766327586390363403000000}{16155146680896107863213416098535762926181898572279}r^{3} + \tfrac{49607330701839093502168550993880180658065000000}{3171065904103137883917151919165713711668601}r^{2} + \tfrac{86943461278789795296227546127346654500000}{6645212181827507092761081352449}r + 1$$

20

$\Delta'(r) = (r-1)r^9(-\frac{81}{26800}r^{45} - \frac{5633195387585655909205183329773}{6524782699847965260642688600000}r^{44} - \frac{18511851290680295821674958382913861338971}{15027264912383421409224514145467323000000}r^{43} - \frac{1129285107246775051349922915235082309059971397953}{964586985869093484764388491547670179462375000000}r^{42} - \frac{27951381645948785379888787463235668294504668152854207}{33521785350500179520029738666566567318112500000000}r^{41} - \frac{60827942349203178778068863211998358870186907289615783102254391}{128107451583325072495694110581784334585477969992402500000000000}r^{40} - \frac{8562912202100990311905987439041329717463210575569086154366093523}{3801449378502983468830969770464264501995093249193750000000000}r^{39} - \frac{44621655705123136077347662614843260751907658760768760237116789155653}{487128584645310881648768554872349322612799806361050625000000000000}r^{38} - \frac{80492686432889156295605195823804626469146709501499514533353144038439997}{24683026806061832059907488479270518517391753824590053828125000000000}r^{37} - \frac{21572467109048293670971951873184793848755659696094963103678001793931386703}{20885638066677040506909517901519772070237196773004554687500000000000000}r^{36} - \frac{239415953278725872564939873433650620858077137653967723511159186173249868847}{81221925814818849086020368072813244693981412157839066015625000000000000}r^{35} - \frac{2864109761580298195855326746967946810954371347588936351918456426483017749497}{374132492926527769411983585217289355480150599309731130859375000000000000}r^{34} - \frac{182389272432036105115132679246396634513515042603013756970025287609794539687}{100031213898250582558572453310727890833640265499654428671875000000000000}r^{33} - \frac{7049970065241562148699334871330114615986042087207199063642535727045336420054}{175980839265440839686377464157762030170293059675317976367187500000000000}r^{32} - \frac{193732072860728817404284511727415205688185855885139395629838358724673859797127811}{237574133008345133576609576612978740729895630516792680957031250000000000000}r^{31} - \frac{1832297077739375209101556520736061152596970134706256943289066632995478163038960819}{118787066504172566788304788306489370364947815280839634047851562500000000000}r^{30} - \frac{80690365377769302841410373659439059544672078289902985512877385073728956242240459}{29696766626043141697076197076622342591236953820209908517169289062500000000}r^{29} - \frac{265118588319449990408878546907749369413286006479697580445186959644220145865382569}{593935332520862833941523941532446851824739076404198170239257812500000000}r^{28} - \frac{20835760488542716315400309426505966512401798476077984767119841711534565417433}{304331109623322516929120505968381398028364839680590820312500000000}r^{27} - \frac{194264688263116858901114319916712721812166672626838952059665585076249286196581011}{19797844417362094464717464717748228394157996921347327234130859375000000000}r^{26} - \frac{6174768978107861419038912615027599943737329385627295727458353708683921405584637}{195942997477312938288591249238312590593933499369727643140885217392697551615218471}r^{25} - \frac{4695140968544370228786750525948196457112561868087890673828125000000000}{118787066504172566788304788306489370364947815280839634047851562500000000}r^{24} - \frac{2001247114991909053457648771421484826742444357097579819825668922409978407402741074}{103293101307976145033308511570860322056476363611137735948242187500000000}r^{23} - \frac{110121595804291167778052089636529865854549031408703460653213521743695293}{516465506539880725166542557854301610282381805568867974121093750000000}r^{22} - \frac{3266158267899045899387602034756597336912684022915956406521959583306986259694336}{148483833130215708485380985383111712956184769101049542559814453125000000000}r^{21} - \frac{28950987185871414926281730160961075903162645894464257554003230301553461254693}{136067659225856319345137214554970641884247211089163383789062500000000}r^{20} - \frac{176309090921596426518287022243129931752659563772107682225665640489658989514832511201}{913746665416712052217729140819149002807290886775689492675781250000000}r^{19} - \frac{54105352084418197792331252632162415905145343333892923367233669020289977649717366648}{329964073622701574411957745295803806569299486891220568847656250000}r^{18} - \frac{1549649686748338374454987051806469178896730084238877170009170310032316093437520631}{11878706650417256678830478830648937036494781528083963404785156250000000}r^{17} - \frac{11354202221936029573206088914280994188843657036730490068063290376818319563443734}{11703159261494834166335447123792056193590917716502728708803240987316670851550989}r^{16} - \frac{168370492982626138238019005126063940078734270642286621093750000000}{11330668168315445571882115643884537226967185502728708803240987316670851550989}r^{15} - \frac{84578352318555854318068425149237494160595130473252346594726562500000000}{1948287133084673885325648478873113066253537031464594726562500000000}r^{14} - \frac{513699704192136461211254644217834570927806693590595958874194982516339931}{197978444173620944647174647177482283941579692134732723413085937500000000}r^{13} - \frac{217833415291497487743317152025992846465590849994367955140829072881017947763137}{152291110902785342036288190136524833801215147795904095537941508958064922999084167}r^{12} - \frac{39661125052776054791853135797201698060179454095537941508958064922999084167}{54841674286321591315006827472986782255282533278319313964843750000000}r^{11} - \frac{1099102683812100152919952408177060036078549127470602338870251321180049}{329964073622701574411957745295803806569299486891220568847656250000}r^{10} - \frac{5489104698106140057625570314166626863533545896970407245157753323808738467272477}{3959568883472418892943492943549645678831593842694654482617187500000000}r^9 - \frac{6131106832511602919256452710496072112438718828437966369670406716840091109299}{118787066504172566788304788306489370364947815280839634047851562500000000}r^8 - \frac{2689414305873757566525592643024364570485198203754824669933483339254158996081}{1583827553388967557177397177419858271532637532677817873046785734468272496}r^7 - \frac{1649226232836823977546802733108121982548768237130739009321510672074046599614}{33939161858335019082372796658996962961413666150881132401367187500000}r^6 - \frac{20625289867463619046424933470290161899000367214515775332380073846727247}{174046983888975337557579315845998100585315974810837128906250000}r^5 - \frac{352291059682303397502433427535140117512763041819680073823551579}{14712340142763950444273705121267946750509856084622460937500000}r^4 - \frac{299629032926479760253255030490759299920442258702724495157042233}{7784306953843360023425240804903675529370294224679687500000}r^3 - \frac{23927768287982533616137767222151734069149516521365218297837}{51843536156132933889459270665118567679183632812500000}r^2 - \frac{3457229600180911245094160397928111958314092757588137}{941979478387786060621522313627702165963122558593757}r - \frac{2739873262687350028588121781119797877}{18869431352372327272248476992968750007})$

$$\Delta''(r) = r^{57} + \frac{6645212181827507092761081352444 9}{245378435104996693552026824850 0} r^{56} + \frac{317976025023101686344677467377844 8377118601}{869434612787897952962275461273466545000 0} r^{55} + \frac{16289490694520104637404694866176922559146 2519982279}{4960733070183909350216855099388018065806500000 0} r^{54} + \frac{4837856349945827907049699704679167465802145415692 2194833}{219393380761953574922690633625533448697835826900000 0} r^{53} + \frac{23165656410740412765753349987634542845001089424410888 3079531}{1960828340559960075871547538028214477369077029187500000} r^{52} + \frac{7168431160720568553056298409926485056517907420131597702553 5267113}{13625974395673031654511008527897883150008386191917500000000} r^{51} + \frac{28592601678155249175725175871296562709509161090205465962716 5571887}{142748303192765093523448660768454013952468807724850000000 0} r^{50} + \frac{1180522227953734487009809473364043434082066606093404727048111920 7967299}{17731480481089316092015175397535153431059129515422247500000000 0} r^{49} + \frac{29408170973460450929880681205282873722784839231121196414033100 46791028651}{149800438547133877329093723184538319277963747349236188750000000 00} r^{48} + \frac{188293921867802609980121265755191801371358726341075995041060481770 01370983}{3620177264889068702119764976959676049217457227606541228125000000 00} r^{47} + \frac{47561407068113171420696418508451533053175046368911395678068968300303 173325997}{3801186128133522137225753225807659851678330088968686828953125000000 00} r^{46} + \frac{52403735730963813999359428464738466770222860649449681668213515681834 6707335427}{1900593064066761068612876612903829925839165044493434144765625000000 00} r^{45} + \frac{8459313447387326074946974058059972437796587091359021440657895512613 663035609}{15084071937037786258832354070665316871739405115027255117187500000 00} r^{44} + \frac{28777512463063488956648718087102988792698777595332316163272713706725 52413629167}{27151329486668015265898237327197570369130929207049059210937500000 000} r^{43} + \frac{2571513294866680152658982373271975703691309292070490592109375000000 0000}{2578756730434961337682727762586159353112520159903506447727051475200 0597178416409} r^{42} + \frac{137724135077301526711078015427813762741968481485031459756250000000 00}{4923690240077702061587827459572665979000056372641367109988420049606 172193809987419} r^{41} + \frac{15838275533889675571773971774198582715326375370778617873046875000000 0000}{8272547394119159712041461652470849424015939526050118626835229135981 00754467256291} r^{40} + \frac{169695809291675095411863983294984814807068307544056620068359537500000 0000}{8616210554080518945933099499459364853363012529112111429131028867463 619051637282857} r^{39} + \frac{118787066504172566788304788306489370364947815280839634047851562500000 0000}{244315156207566683731121214566700356307778212944866802042856090003 5345241377758078 33} r^{38} + \frac{23757413300834513357660957661297874072989563056167926809570312500000 00000}{552196271145602876042689106655107647236337977956942996053315489396 1457922388134417} r^{37} + \frac{3959568883472418892943492943549645678831593842694654468261718750000 0000}{21561564951167919750832487402775766820265336541427813068300970403 45127109261937089 9} r^{36} + \frac{118787066504172566788304788306489370364947815280839634047851562500 000000}{180094645405691272853633604334198126975439052098418486917560976 9133977715249645226 1} r^{35} + \frac{79191377669448377858869858870992913576631876853893089365234375000 00000}{65317677663253357151894511940209678389193319191213490439059004 03145987635512480225 9} r^{34} + \frac{23757413300834513357660957661297874072989563056167926809570312500000 00000}{63617234024657046013910089740208352718700135368820772582183306389360 49738047840343} r^{33} + \frac{1979784441736209446471746471774822839415796921347327234130859375 0000000}{332218137183117318606269693439063144199440187806610451102175232 25239389 76696259767} r^{32} + \frac{91374666541671205221772914081914900280729088677568949267578125000 000000}{41174616924114353854336589212302557533863489796445342063734864986 04632172945064269} r^{31} + \frac{10329310130797614503330851157086032205647636111377359482421875000 00000000}{10066712375173857530179267712688226692532522009297664423127732 132483154737817125854 9} r^{30} + \frac{2375741330083451335766095766129787407298956305616792680957031250 0000000000}{5823928141793754818105611918824670762640965875785940282529741245637 3577529537229} r^{29} + \frac{1333188176253339694593768667861833561896159542994833154296875000 0000000}{5823928141793754818105611918824670762640965875785940282 5297414256373577529537229} r^{28} + \frac{1333188176253339694593768667861833561896159542994833154296875000 0000000}{1006671237517385753017926771268822669253252200929766442312773 21324831547378171258549} r^{27} + \frac{23757413300834513357660957661297874072989563056167926809570312500000 00000}{41174616924114353854336589212302557533863489796445342063734864986 04632172945064269} r^{26} + \frac{103293101307976145033308511570860322056476361113773594824218750 00000000}{33221813718311731860626969343906314419944018780661045110217523 225239389 76696259767} r^{25} + \frac{91374666541671205221772914081914900280729088677568949267578125000 000000}{63617234024657046013910089740208352718700135368820772582183306 389360 49738047840343} r^{24} + \frac{1979784441736209446471746471774822839415796921347327234130859375 0000000}{65317677663253357151894511940209678389193319191213490439059004 03145987635512480225 9} r^{23} + \frac{23757413300834513357660957661297874072989563056167926809570312500 0000000}{180094645405691272853633604334198126975439052098418486917560976 9133977715249645226 1} r^{22} + \frac{79191377669448377858869858870992913576631876853893089365234375000 0000000}{21561564951167919750832487402775766820265336541427813068300970403 45127109261937089 9} r^{21} + \frac{118787066504172566788304788306489370364947815280839634047851562500 00000}{552196271145602876042689106655107647236337977956942996053315489396 1457922388134417} r^{20} + \frac{3959568883472418892943492943549645678831593842694654468261718750 0000}{24431515620756668373112121456670035630777821294486680204285609000 3534524137775807833} r^{19} + \frac{237574133008345133576609576612978740729895630561679268095703125 00000000}{86162105540805189459330994945936485336301252911211142913102886 7463619051637282857} r^{18} + \frac{1187870665041725667883047883064893703649478152808396340478515625000 0000}{8272547394119159712041461652470849424015939526050118626835229135981 00754467256291} r^{17} + \frac{1696958092916750954118639832949848148070683075440566200683593750 00000000}{49236902400777020615878274595726659790005637264136710998842004 9606172193809987419} r^{16} + \frac{158382755338896755717739717741985827153263753707786178730468750 00000000}{25787567304349613376827277625861593531125201599035064477270 514752005971784 16409} r^{15} + \frac{137724135077301526711078015427813762741968481485031459756250000 00000}{28777512463063488956648718087102988792698777595332316163272713 70672552413629167} r^{14} + \frac{2715132948666801526589823732719757036913092920704905921093750000 00000}{84593134473873260749469740580599724377965870913590214406578955 12613663035609} r^{13} + \frac{15084071937037786258832354070665316871739405115027255117187500 0000}{524037357309638139993594284647384667702228606494496816682135156 81834 6707335427} r^{12} + \frac{1900593064066761068612876612903829925839165044493434144765625 000000000}{475614070681131714206964185084515330531750463689113956780 68968300303 173325997} r^{11} + \frac{38011861281335221372257532258076598516783300889686682895312500 0000}{188293921867802609980121265755191801371358726341075995041060481770 01370983} r^{10} + \frac{362017726488906870211976497695967604921745722760654122812500000 00}{29408170973460450929880681205282873722784839231121196414033100 46791028651} r^{9} +$$

$$\frac{1180522227953734487009809473364043434082066606093404727048111920796 7299}{17731480481089316092015175397353515343105912951542242750000000 00}r^8 +$$
$$\frac{28592601678155249175725175871296562709509161090205465962716557188 7}{1427483031927509352344866076845401395246880772485000000000 0}r^7 +$$
$$\frac{716843116072056855305629840992648505651790742013159770253526711 3}{1362597439567303165451100852789788315000838619191750000000 0}r^6 +$$
$$\frac{23165656410740412765753349987634542845001089424410888307953 1}{1960828340559960075871547538028214477369077029187500000}r^5 +$$
$$\frac{48378563499458279070496997046791674658021454156922194833}{21939338076195357492269063362553448697835826900000 00}r^4 +$$
$$\frac{16289490694520104637404694866176922559146251998227 9}{49607330701839093502168550993880180658065000000}r^3 +$$
$$\frac{3179760250231016863446774673778448377118601}{8694346127878979529622754612734665450000}r^2 +$$
$$\frac{6645212181827507092761081352444 9}{24537843510499669355202682485 00}r + 1$$

%


\begin{thebibliography}{10}

\bibitem{M}
P.~A.~P. Moran, ``Random processes in genetics,'' \emph{Proc. Cambridge Philos.
  Soc.}, vol.~54, pp. 60--71, 1958.

\bibitem{LHN}
E.~Lieberman, C.~Hauert, and M.~A. Nowak, ``Evolutionary dynamics on graphs,''
  \emph{Nature}, vol. 433, no. 7023, pp. 312--316, Jan. 2005.

\bibitem{N}
M.~A. Nowak, \emph{Evolutionary Dynamics: Exploring the Equations of
  Life}.\hskip 1em plus 0.5em minus 0.4em\relax Belknap Press of Harvard
  University Press, Sep. 2006.

\bibitem{BR}
M.~Broom and J.~Rycht{\'a}{\v r}, ``An analysis of the fixation probability of
  a mutant on special classes of non-directed graphs,'' \emph{Proceedings of
  the Royal Society of London A: Mathematical, Physical and Engineering
  Sciences}, vol. 464, no. 2098, pp. 2609--2627, 2008. [Online]. Available:
  \url{http://rspa.royalsocietypublishing.org/content/464/2098/2609}

\bibitem{BHRS}
M.~Broom, C.~Hadjichrysanthou, J.~Rycht{\'a}{\v r}, and B.~T. Stadler, ``Two
  results on evolutionary processes on general non-directed graphs,''
  \emph{Proceedings of the Royal Society of London A: Mathematical, Physical
  and Engineering Sciences}, vol. 466, no. 2121, pp. 2795--2798, 2010.
  [Online]. Available:
  \url{http://rspa.royalsocietypublishing.org/content/466/2121/2795}


\bibitem{SR2}
P.~Shakarian, P.~Roos, and A.~Johnson, ``A review of evolutionary graph theory
  with applications to game theory,'' \emph{Biosystems}, vol. 107, no.~2, pp.
  66 -- 80, 2012.

\bibitem{ShakarianBio}
P.~Shakarian, P.~Roos, and G.~Moores, ``A novel analytical method for
  evolutionary graph theory problems,'' \emph{Biosystems}, vol. 111, no.~2, pp.
  136 -- 144, 2013. [Online]. Available:
  \url{http://www.sciencedirect.com/science/article/pii/S030326471300021X}


\bibitem{Voorhees1}
B.~Voorhees and A.~Murray, ``Fixation probabilities for simple digraphs,''
  \emph{Proceedings of the Royal Society of London A: Mathematical, Physical
  and Engineering Sciences}, vol. 469, no. 2154, 2013. [Online]. Available:
  \url{http://rspa.royalsocietypublishing.org/content/469/2154/20120676}


\bibitem{AGSL}
F.~Alcalde~Cuesta, P.~Gonz{\'a}lez~Sequeiros, and {\'A}.~Lozano~Rojo, ``Fast
  and asymptotic computation of the fixation probability for moran processes on
  graphs,'' \emph{Biosystems}, vol. 129, pp. 25 -- 35, 2015. [Online].
  Available:
  \url{http://www.sciencedirect.com/science/article/pii/S0303264715000088}


\bibitem{Adlam&al}
B.~Adlam, K.~Chatterjee, and M.~A. Nowak, ``Amplifiers of selection,''
  \emph{Proceedings of the Royal Society of London A: Mathematical, Physical
  and Engineering Sciences}, vol. 471, no. 2181, 2015. [Online]. Available:
  \url{http://rspa.royalsocietypublishing.org/content/471/2181/20150114}


\bibitem{HindersinTraulsen}
L.~Hindersin and A.~Traulsen, ``Most undirected random graphs are amplifiers of
  selection for birth-death dynamics, but suppressors of selection for
  death-birth dynamics,'' \emph{PLoS Comput Biol}, vol.~11, no.~11, pp. 1--14,
  11 2015. [Online]. Available:
  \url{http://dx.doi.org/10.1371%2Fjournal.pcbi.1004437}


\bibitem{KKK}
K.~Kaveh, N.~L. Komarova, and M.~Kohandel, ``The duality of spatial
  death{\textendash}birth and birth{\textendash}death processes and limitations
  of the isothermal theorem,'' \emph{Royal Society Open Science}, vol.~2,
  no.~4, 2015. [Online]. Available:
  \url{http://rsos.royalsocietypublishing.org/content/2/4/140465}


\bibitem{Monk&al}
T.~Monk, P.~Green, and M.~Paulin, ``Martingales and fixation probabilities of
  evolutionary graphs,'' \emph{Proceedings of the Royal Society of London A:
  Mathematical, Physical and Engineering Sciences}, vol. 470, no. 2165, 2014.
  [Online]. Available:
  \url{http://rspa.royalsocietypublishing.org/content/470/2165/20130730}


\bibitem{BRS2}
M.~Broom, J.~Rycht{\'a}{\v{r}}, and B.~T. Stadler, ``Evolutionary dynamics on
  graphs---the effect of graph structure and initial placement on mutant
  spread,'' \emph{J. Stat. Theory Pract.}, vol.~5, no.~3, pp. 369--381, 2011.
  [Online]. Available: \url{http://dx.doi.org/10.1080/15598608.2011.10412035}


\bibitem{SciRep}

F.~Alcalde~Cuesta, P.~Gonz{\'a}lez~Sequeiros, and {\'A}.~Lozano~Rojo,
  ``Exploring the topological sources of robustness against invasion in
  biological and technological networks,'' \emph{Scientific Reports}, vol.~6,
  pp. 20\,666 EP --, 02 2016. [Online]. Available:
  \url{http://dx.doi.org/10.1038/srep20666}


\bibitem{AGSL2}
F.~Alcalde~Cuesta, P.~Gonz\'alez~Sequeiros, {\'A}.~Lozano~Rojo, and R.~Vigara,
  ``An accurate database of the fixation probabilities of all undirected graphs
  of order 10 or less,'' Submitted for publication, 2016.

\bibitem{sage}
W.~A. Stein \emph{et~al.}, \emph{{S}age {M}athematics {S}oftware ({V}ersion
  5.8)}, The Sage Development Team, 2013, {\tt http://www.sagemath.org}.

\bibitem{KT}
S.~Karlin and H.~M. Taylor, \emph{A first course in stochastic processes},
  2nd~ed.\hskip 1em plus 0.5em minus 0.4em\relax Academic Press [A subsidiary
  of Harcourt Brace Jovanovich Publishers], New York-London, 1975.

\bibitem{TanLu}
S.~Tan and J.~Lu, ``Characterizing the effect of population heterogeneity on
  evolutionary dynamics on complex networks,'' \emph{Sci. Rep.}, vol.~4, p.
  2014/05/22/online, 2014.

\bibitem{HWDT}
L.~Hindersin, B.~Werner, D.~Dingli, and A.~Traulsen, ``Should tissue structure
  suppress or amplify selection to minimize cancer risk?'' \emph{Biology
  Direct}, vol.~11, no.~1, p.~41, 2016. [Online]. Available:
  \url{http://dx.doi.org/10.1186/s13062-016-0140-7}


\bibitem{KSN}
N.~L. Komarova, A.~Sengupta, and M.~A. Nowak, ``Mutation-selection networks of
  cancer initiation: tumor suppressor genes and chromosomal instability,''
  \emph{Journal of Theoretical Biology}, vol. 223, no.~4, pp. 433 -- 450, 2003.
  [Online]. Available:
  \url{http://www.sciencedirect.com/science/article/pii/S0022519303001206}


\bibitem{SSL}
V.~Salas-Fum\'as, C.~S\'aenz-Royo, and A.~Lozano-Rojo, ``Organisational
  structure and performance of consensus decisions through mutual influences: A
  computer simulation approach,'' \emph{Decision Support Systems}, vol.~86, pp.
  61 -- 72, 2016. [Online]. Available:
  \url{http://www.sciencedirect.com/science/article/pii/S0167923616300422}


\bibitem{KT:ISM}
H.~M. Taylor and S.~Karlin, \emph{An introduction to stochastic modeling},
  3rd~ed.\hskip 1em plus 0.5em minus 0.4em\relax San Diego, CA: Academic Press
  Inc., 1998.

\bibitem{FangHavas}

X.~G. Fang and G.~Havas, ``On the worst-case complexity of integer gaussian
  elimination,'' in \emph{Proceedings of the 1997 International Symposium on
  Symbolic and Algebraic Computation}, ser. ISSAC '97.\hskip 1em plus 0.5em
  minus 0.4em\relax New York, NY, USA: ACM, 1997, pp. 28--31. [Online].
  Available: \url{http://doi.acm.org/10.1145/258726.258740}


\end{thebibliography}
\end{document}